\documentclass[aps,pra,twocolumn,footinbib, superscriptaddress]{revtex4-1}

\usepackage{graphicx}
\usepackage{hyperref}
\usepackage{amsmath,amssymb}
\usepackage{bbold}
\usepackage[utf8]{inputenc}
\usepackage{orcidlink}

\begin{document}
	
\title{Security Analysis of Ensemble-Based Quantum Token Protocol Under Advanced Attacks}
	
\renewcommand{\figurename}{FIG.}
\renewcommand{\tablename}{TAB.}
	
\author{Bernd Bauerhenne \orcidlink{0000-0002-3397-2290}}
\email{bauerhenne@uni-kassel.de}
\affiliation{Experimental Physics I, University of Kassel, Heinrich-Plett-Strasse 40, 34132 Kassel, Germany}
	
\author{Lucas Tsunaki \orcidlink{0009-0003-3534-6300}}
\affiliation{Department Spins in Energy Conversion and Quantum Information Science (ASPIN), Helmholtz-Zentrum Berlin für Materialien und Energie GmbH, Hahn-Meitner-Platz 1, 14109 Berlin, Germany}
	
\author{Jan Thieme}
\affiliation{Experimental Physics I, University of Kassel, Heinrich-Plett-Strasse 40, 34132 Kassel, Germany}

\author{Boris Naydenov \orcidlink{0000-0002-5215-3880}}
\affiliation{Department Spins in Energy Conversion and Quantum Information Science (ASPIN), Helmholtz-Zentrum Berlin für Materialien und Energie GmbH, Hahn-Meitner-Platz 1, 14109 Berlin, Germany}
\affiliation{Berlin Joint EPR Laboratory, Fachbereich Physik, Freie Universität Berlin, 14195 Berlin, Germany}
	
\author{Kilian Singer \orcidlink{0000-0001-9726-0367}}
\email{ks@uni-kassel.de}
\affiliation{Experimental Physics I, University of Kassel, Heinrich-Plett-Strasse 40, 34132 Kassel, Germany}



\date{\today}
	
\begin{abstract}
We present and characterize advanced attacks on an ensemble-based quantum token protocol that allows for implementing non-clonable quantum coins. 
Multiple differently initialized tokens of identically prepared qubit ensembles are combined to a quantum coin that can be issued by a bank. 
A sophisticated attempt to copy tokens can assume that measurements on sub-ensembles can be carried through and that even individual qubits can be measured. 
Even though such an advanced attack might be perceived as technically unfeasible, we prove the security of the protocol under these conditions. 
We performed numerical simulations and verified our results by experiments on the IBM Quantum Platforms for different types of advanced attacks. 
Finally, we demonstrate that the security of the quantum coin can be made high by increasing the number of tokens.
This paper in conjunction with provided numerical simulation tools verified against experimental data from the IBM Quantum Platforms allows for securely implementing our ensemble-based quantum token protocol with arbitrary quantum systems.
\end{abstract}

	
\maketitle

\section{Introduction}

Quantum tokens \cite{Wiesner1983,Gavinsky2011,Molina2013,Pastawski2012,Georgiou2015, Moulick2016,Amiri2017,Bozzio2019,Kumar2019,Horodecki2020, Kent2022} are proposed as an alternative to classical identification tokens due to improved security guaranteed by the laws of quantum physics. 
The security is based on the quantum no-cloning theorem, the fact that quantum states cannot be cloned with arbitrary precision.
The theorem is a direct consequence of the linearity of quantum mechanics, but its necessity also follows from the fact that measurements on clones of an unknown state could easily violate the Heisenberg-uncertainty relation. 

Despite the great potential of the quantum token application, the experimental implementation with single qubits \cite{Pastawski2012} faces many technical challenges. 
As an example, single qubit control poses typically higher demands on readout, requiring highly sensitive detection techniques to accurately measure the quantum state population. 
The use of ensembles in a redundant quantum parallelism regime reduces errors and decoherence. 
To simplify quantum token implementations, we have successfully designed a patented ensemble-based quantum token \cite{patent_quantum_token} that is technologically less demanding than conventional single-qubit-based methods. 
Such a token consists of an ensemble of identical qubits.
Using such an implementation would typically render the quantum-no-cloning theorem inapplicable as a protection scheme, because an ensemble already consists of identical qubit copies.

However, the quantum projection noise will be reduced when a measurement of the token is performed in the proper basis, which can be understood by the fact that measurements in the Eigenstate basis are free of quantum projection noise \cite{OURPRX,Wineland1993}. 
Thus, the resulting noise  reveals a copy operation of a forger, as the cloned token will show increased quantum projection noise when the cloning operation is performed in the wrong basis.

Combining differently initialized tokens to a coin allows to define a ensemble-based quantum token protocol with security against unauthorized coin copying that can be made arbitrarily high.
The number of qubits in an ensemble and also the amount of tokens in a coin are crucial design parameters for the security of the protocol.
Other important parameters are intrinsically linked to fundamental characteristics of the underlying quantum platform, such as coherence times and qubit lifetimes.
We present an optimal copying procedure for the individual tokens of the coin that is more efficient than the tomography methods from the literature based on direct inversion \cite{Schmied2016} or on the maximum likelihood method \cite{Paris2016} or on Bayesian experimental design \cite{Hannemann2002}.

Even when using the advanced copying procedure for the individual tokens, the quantum coins can be designed in such a way that the acceptance probabilities of forged coins becomes negligible. 
For this, numerical simulations of the attack scenarios were performed using C++ programs fully parallelized using the message passing interface (MPI). This library called ``DIQTOK-forge" is available on GitHub \cite{githubBauerhenne}.
These findings are further supported by experimental measurements \cite{OURPRX}  performed on five IBM superconducting quantum processors of the Eagle family \cite{hardware1, hardware2, hardware3}: Kyiv, Sherbrooke, Osaka, Brisbane and Kyoto. The experiments were controlled by the Qiskit software \cite{qiskit},  the implementations can be accessed through the author's GitHub repository \cite{github_lucas}. 
The detailed analysis presented in this paper combined with the open source tools allow for designing secure quantum coins for any quantum platform, since the protocol is hardware agnostic.

This work is divided as follows. 
In Sec.~\ref{sec:protocol}, we discuss the framework of the ensemble-based quantum token protocol.
Then we focus on the individual tokens in the coin.
We identified relevant parameters  and benchmarked these for the IBMQ hardware.
Then in Sec.~\ref{sec:attack}, different attack scenarios are considered and tested on IBMQ, where an attacker attempts to read a token and use different methods to create a forged token.
Finally, the safety of the coin as a set of tokens is studied in Sec.~\ref{sec:coin} and the paper is concluded with a final discussion in Sec.~\ref{sec:discussion}.

\section{Ensemble-Based Quantum Token Protocol}
\label{sec:protocol}

In this section, we discuss in detail the realization of the protocol. In Secs.~\ref{sec:description_single_qubit} and ~\ref{sec:description_token}, we describe the mathematical framework for a single qubit and a token composed of identical qubits, respectively. In Sec.~\ref{sec:bank}, we discuss how the bank should generate and accepts its own tokes. 

\begin{figure}[htb!]
    \centering
    \includegraphics[width=0.9\columnwidth]{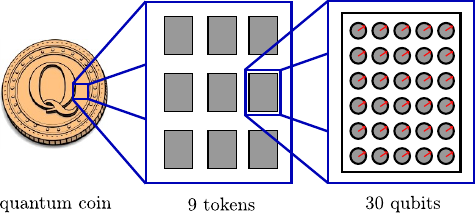}
    \caption{The proposed quantum coin consists of $M$ quantum tokens, each comprising $N$ qubits prepared in the same state. Each quantum token exhibits a distinct qubit state.
    }
    \label{fig:quantum_coin}
\end{figure}

A quantum coin contains $M$ quantum tokens, each with $N$ identical qubits prepared in the same quantum state, but different for each token, as can be seen in Fig.~\ref{fig:quantum_coin}.

\begin{figure}[htb!]
    \centering
    \includegraphics[width=\columnwidth]{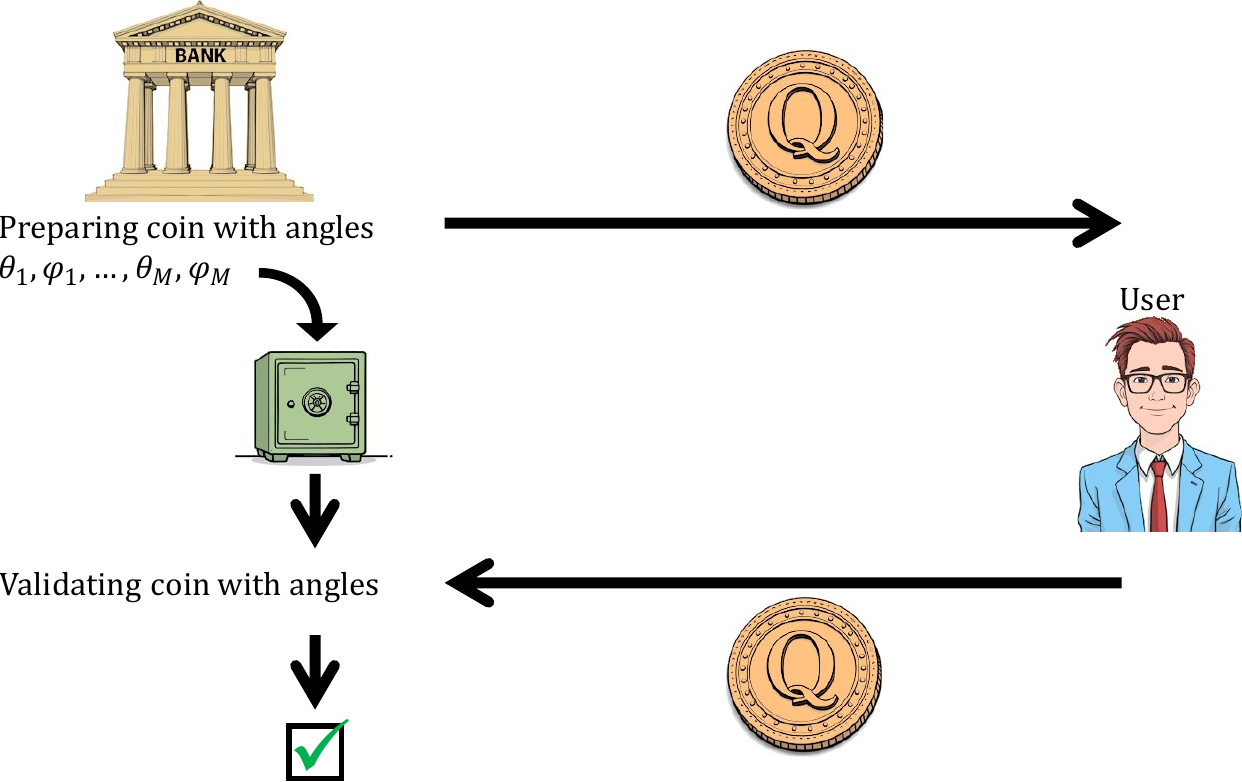}
    \caption{Normal operation of the protocol: 
    The bank selects $M$ secret angle pairs $(\theta_i, \phi_i)$ and prepares a quantum coin by initializing all qubits of the $i$-th quantum token in the state defined by $(\theta_i, \phi_i)$. The coin is provided to the user, while the corresponding angles are securely stored by the bank. Upon return of the coin, the bank verifies its authenticity by measuring each quantum token using the stored angles. The coin is accepted if the number of valid tokens exceeds a predefined threshold. This threshold is chosen such that the probability of the bank accepting its own coin is very high.
    }
    \label{fig:protocol_user}
\end{figure}

\begin{figure}[htb!]
    \centering
    \includegraphics[width=\columnwidth]{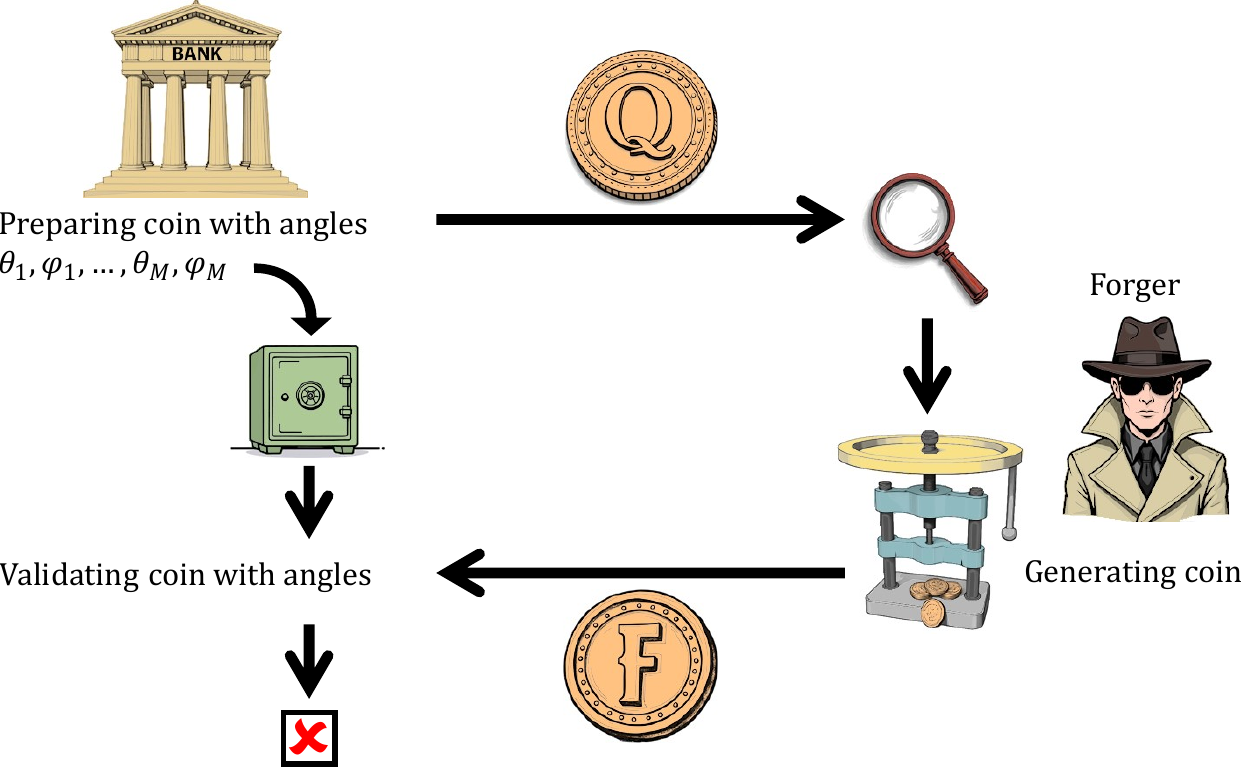}
    \caption{Attack on the protocol: 
    A forger obtains the bank's coin and performs measurements on it to extract information, then uses this information to prepare a forged coin (symbolized by the coin press), which is returned to the bank. The bank attempts to verify the forged coin using the stored secret angles. By design, a larger fraction of quantum tokens in a forged coin are likely to be invalid, resulting in a high probability of rejecting the coin.
    }
    \label{fig:protocol_forger}
\end{figure}

In order to prepare a coin, a bank randomly selects $M$ secret angle pairs $(\theta_i, \phi_i)$, and initializes the qubits in the $i$-th quantum token with these angles.
The prepared quantum coin is then issued to the user, while the set of secret angles is securely kept by the bank.
Upon the coin’s return for verification, the bank measures each quantum token using the stored angles.
The coin is accepted if the number of correctly measured tokens exceeds a predetermined threshold, which is set to ensure that the probability of falsely rejecting a legitimately issued coin remains negligibly small.
This normal operation of the protocol is visualized in Fig.~\ref{fig:protocol_user}.
If a forger copies the coin by using sophisticated methods and provides a forged coin to the bank, the bank will again check the
coin using the stored angles.
Now, by construction, a high number of tokes are not accepted so that the whole coin is rejected with a very high probability.
This procedure is shown in Fig.~\ref{fig:protocol_forger}.

\begin{figure*}[t!]
	\includegraphics[width=0.98\textwidth]{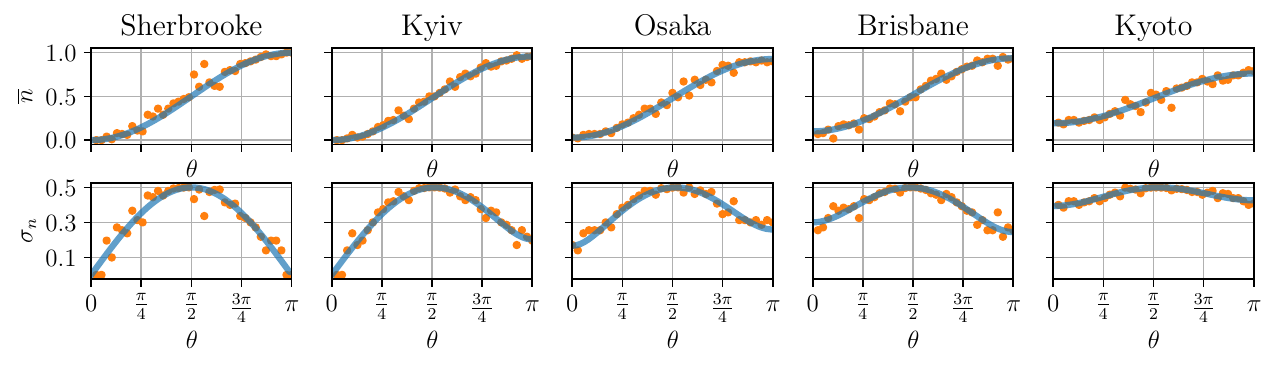}
	\caption{Experimental averaged normalized photon counts $\overline{n}$ and corresponding standard deviations $\sigma_n$ for 100 qubits as a function of $\theta$ for the five IBMQs (data taken from Ref. \cite{OURPRX}).
	Points correspond to the experimental data and lines correspond to the predictions of our model with parameters $P_0$, $P_1$ fitted to the experimental data of $\sigma_n$. $\overline{n}$ curves are related through Eq.~\eqref{equ:mean_n_sigma_n}.}
	\label{fig:fit_params}
\end{figure*}

\subsection{Description of a Single Qubit}\label{sec:description_single_qubit}
In general, the state of a single qubit or an ensemble can be described by a density matrix \cite{Bagan2005,Schmied2016}
\begin{align*}
    \hat{\rho} =& \frac{1}{2}\bigl(\hat{\mathbb{1}} + x\,\hat{\sigma}_x + y\,\hat{\sigma}_y + z\,\hat{\sigma}_z  \bigr)
\end{align*}
in terms of the Pauli matrices $\hat{\mathbb{1}},\hat{\sigma}_x,\hat{\sigma}_y,\hat{\sigma}_z$ and the Bloch vector $\mathbf{r}=(x,y,z) \in \mathbb{R}^3$.
The eigenvectors of $\hat{\rho}$ are 
\begin{align*}
    \lambda^\pm = \frac{1}{2}\left(1\pm \sqrt{x^2+y^2+z^2} \right)
\end{align*}
and must be both non-negative, so that $\mathbf{r}$ must fulfill $||\mathbf{r}||^2 \leq 1$ for a physical state.
Mixed states obey $||\mathbf{r}||^2 < 1$ and pure states have $||\mathbf{r}||^2 = 1$.
In order to simplify the model, we only consider pure states, which are represented by the surface of the Bloch sphere, in contrast to mixed states that are located inside the Bloch sphere.
Thus, we describe the state $|\theta, \phi \rangle $ of a qubit with the polar angle $\theta\in [0,\pi]$ and the azimuthal angle $\phi \in (-\pi,\pi]$.
In the orthonormal basis $|0\rangle$ and $|1 \rangle$, we can represent a general state $|\theta_2, \phi_2\rangle$ as
\begin{align*}
	| \theta_2, \phi_2 \rangle =& e^{-i\, \frac{\phi_2}{2}}\, \cos\left(\frac{\theta_2}{2}\right) |0\rangle + e^{ i\, \frac{\phi_2}{2}}\, \sin\left(\frac{\theta_2}{2} \right) | 1 \rangle
	\\=& e^{-i\, \frac{\phi_2}{2} \, \hat{\sigma}_z} \, e^{-i\, \frac{\theta_2}{2} \,  \hat{\sigma}_y}\,  |0 \rangle.
\end{align*}
Physically, this can be achieved by initializing the qubit in the state $|0\rangle$, performing a rotation around the $y$-axis by $\theta_2$, followed by a rotation around the $z$-axis by $\phi_2$.
We study here only Stern-Gerlach like measurements on the qubit \cite{gerlach}.
Such a measurement uses angles $\theta_1,\phi_1$ for a back rotation around the $z$-axis with $\phi_1$, followed by a back rotation around the $y$-axis by $\theta_1$. This leads to the final state
\begin{align*}
    | \Psi  \rangle =& e^{i\, \frac{\theta_1}{2} \, \hat{\sigma}_y} \,  e^{i\, \frac{\phi_1}{2} \, \hat{\sigma}_z}  \,  e^{-i\, \frac{\phi_2}{2} \, \hat{\sigma}_z} \,  e^{-i\, \frac{\theta_2}{2} \,  \hat{\sigma}_y}\,  |0 \rangle.
\end{align*}
Then, the qubit is measured in the orthonormal basis.
As a concrete example, we take the readout process to yield photons, which are detected with different probabilities. A photon is measured with a probability $P_0 \in [0,1]$ or $P_1 \in [0,1]$ for state $|0\rangle$ or $|1\rangle$ respectively.
Note that $P_0$ and $P_1$ are model parameters to be determined for the considered quantum system, which not necessarily add up to 1 in this definition. Furthermore, they should fulfill $P_0 \neq P_1$, otherwise a measurement would not provide any information about the qubit.
We assign $|0\rangle$ to the dark state, so that $P_0 < P_1$. 
An important quality factor of the hardware can be defined from these probabilities described by the normalized contrast
\begin{align}
    c &\equiv \frac{P_1 -P_0}{P_1 + P_0}.
    \label{equ:contrast_C}{}
\end{align}
This value represents how well we can distinguish between dark and bright states upon measurement.

The probability to measure a photon from the qubit in an arbitrary state is given by
\begin{align}
	p_\text{q} =& P_0 \, \Bigr| \langle 0 | \Psi \rangle \Bigl|^2 + P_1 \, \Bigr| \langle 1 | \Psi \rangle \Bigl|^2 = P_1 - (P_1-P_0) \, \Bigr| \langle 0 | \Psi \rangle \Bigl|^2,
	\label{equ:p_qubit}
\end{align}
where we used $| \langle 0 | \Psi \rangle |^2+| \langle 1 | \Psi \rangle |^2 = 1$.
We further obtain \cite{OURPRX}
\begin{align*}
	  & \Bigr| \langle 0 | \Psi \rangle \Bigl|^2 =  \Bigr| \langle 0 | e^{i\, \frac{\theta_1}{2} \, \hat{\sigma}_y} \,  e^{i\, \frac{\phi_1}{2} \, \hat{\sigma}_z}  \,  e^{-i\, \frac{\phi_2}{2} \, \hat{\sigma}_z} \,  e^{-i\, \frac{\theta_2}{2} \,  \hat{\sigma}_y}\,  |0 \rangle \Bigl|^2 \nonumber \\
	  =& \Bigr| \langle \theta_1,\phi_1 | \theta_2,\phi_2 \rangle\bigr. \Bigl|^2 \nonumber \\ 
	 =&\frac{1}{2} \Bigl( 1+ \cos(\theta_1) \, \cos(\theta_2) + \sin(\theta_1) \, \sin(\theta_2) \, \cos(\phi_1 - \phi_2) \Bigr).
\end{align*}
Substituting the above result in Eq.~\eqref{equ:p_qubit}, we get
\begin{align}
    & p_\text{q}(P_0, P_1, \theta_1, \phi_1, \theta_2, \phi_2) \nonumber \\
    =& \frac{P_0+P_1}{2} - \frac{P_1-P_0}{2} \Bigl(\cos(\theta_1) \, \cos(\theta_2) \nonumber \\
    & \qquad + \sin(\theta_1) \, \sin(\theta_2) \, \cos(\phi_1 - \phi_2) \Bigr).
    \label{equ:p_qubit_fin}
\end{align}
If we rotate the coordinate system around the $z$-axis, an offset angle $\Delta \phi$ is added to all angles $\phi$. Thus, $p_\text{q}$ will not change, since $\phi$ angles only occur in the term $\cos(\phi_1-\phi_2)$ in Eq. \eqref{equ:p_qubit_fin} and thus the added value of $\Delta \phi$ is canceled.

\subsection{Description of a Token of Identical Qubits}\label{sec:description_token}

Now, we consider a quantum token consisting of $N$ such identical qubits. 
After the measurement using the angles $\theta_1,\phi_1$ the probability to measure $n\in \{0,1,\ldots, N\}$ photons from these qubits  is given by the binomial distribution with probability $p_q$
\begin{align}
	p_\text{t}(N, n, P_0, P_1, \theta_1,\phi_1,\theta_2,\phi_2) = \binom{N}{n} \, p_q^n \, (1-p_q)^{N-n}.
	\label{equ:p_token}
\end{align}
	
In order to determine the parameters $P_0$ and $P_1$, one can drive a Rabi oscillation on the quantum token \cite{rabi}. This is modelled by preparing the state with different $\theta_2$ settings and keeping $\phi_2=0$ fixed. The subsequent measurement is performed using $\theta_1=\pi$, $\phi_1=0$.
Averaging over many measurements gives the averaged normalized photon counts $\overline{n}$ and the corresponding standard deviation $\sigma_n$ as a function of the angle $\theta_2$.
From Eq.~\eqref{equ:p_token}, we obtain
\begin{align}
	\overline{n} = p_\text{q}, \qquad \sigma_n =  p_\text{q} \, (1 - p_\text{q}).
	\label{equ:mean_n_sigma_n}
\end{align}

The experimental values of $\overline{n}$ and $\sigma_n$ as a function of $\theta$ for 100 qubits were measured with five IBMQ, as shown in Fig.~\ref{fig:fit_params}. In the graphs $\overline{n}$ represents the Rabi oscillations for $\theta$ from $0$ to $\pi$. The standard deviation $\sigma_n$ at $\theta\in\{0,\pi\}$ shows a minimum and ideally a zero value due to the fact that the quantum projection noise vanishes at the poles of the Bloch sphere. One can clearly identify that for the superposition state at $\theta=\pi/2$ the standard deviation becomes maximal due to the fact that this state has the largest projection noise. 
Depending on the implementation of the qubit, background photons collected during the readout can lead to reduction of the amplitude of the cosine approaching 0.5. Additionally, increased relative shot noise due to low photon counts during read out leads to a reduction of the amplitude of the cosine for the bright state. If the state at $\theta=0$ is a dark state then this reduction is only observed at $\theta=\pi$. 
The fit of the experimental curves permits us to obtain the values of $P_0$ and $P_1$ for each hardware, as presented in Tab.~\ref{tab:fit_params}. 
Comparing the different IBMQ platforms Sherbrooke has the best parameters, given by the hardware's longer coherence times and smaller gate and readout errors. Inversely, Kyoto has the worst parameters of the three. Note, in our companion study Ref.~\cite{OURPRX}, we used a different model focusing on the noise description in order to model the quantum token. Here, instead, we use a statistical description in order to calculate the probability of acceptance for forged quantum tokens, both representations being in good agreement. The statistical description is better suited for the following derivations.

\begin{table}[tb!]
	\begin{tabular}{|l|c|c|c|}
		\hline
		IBMQ       & $P_0$ & $P_1$ & $c$ \\ 
		\hline
		Sherbrooke & $2.104\cdot10^{-34}$ & 0.999999991 & $\sim 1$\\
		Kyiv       & $7.197\cdot10^{-13}$ & 0.9571 & $\sim 1$\\
		Osaka      &     0.02855 &      0.9274 & 0.9403\\ 
		Brisbane   &     0.1003 &      0.9362  & 0.8065\\ 
		Kyoto      &     0.1916 &      0.7615  & 0.5979\\
		\hline
	\end{tabular}
	\caption{Fitted values of $P_0$ and $P_1$ from the experimental curves of the average normalized photon counts $\overline{n}$ and their corresponding standard deviation $ \sigma_n$ measured in the IBMQ hardware (Fig.~\ref{fig:fit_params}). Additionally, the normalized contrast $c$ is included. Sherbrooke has the best performance among the five hardware platforms, while Kyoto has the worst specifications due to dissimilar coherence times and average errors in readout and gate operations of the qubits. These two parameters are the basis to describe the following development of the quantum coin model.}
	\label{tab:fit_params}
\end{table}


\subsection{Bank Generates Quantum Tokens}
\label{sec:bank}

A bank generates a quantum token with the angles $\theta_\text{b}$, $\phi_\text{b}$.
In order to derive the safety of the token with statistical methods, we consider both angles as independent random variables with a uniform probability distribution on the unit sphere, so that corresponding probability densities are given by
\begin{align}
	f_{\theta}(\theta) = \frac{\sin(\theta)}{2}, \qquad f_{\phi}(\phi) = \frac{1}{2\pi}.
	\label{equ:fbtheta_fbphi}
\end{align}
For verification, the bank measures the token with preparation angles $\theta_\text{b}$, $\phi_\text{b}$.
We denote  the $P_0$ and $P_1$ hardware quality parameters of the bank setup by $P_{0\text{b}}, P_{1\text{b}}$.
The bank sets a threshold $n_\text{T}$ for maximum photon counts to accept the token.
This is because for measuring the token the bank projects it into the dark state $|0\rangle$.
Thus, the average probability that the bank accepts a not forged token is given by
\begin{align}
	\overline{p}_\text{b} =& \int \limits_{0}^{\pi} d\theta_\text{b} \int \limits_{-\pi}^{\pi} d\phi_\text{b} \, f_{\theta}(\theta_\text{b}) \, f_{\phi}(\phi_\text{b}) \,  \times \nonumber \\ 
	& \times \sum \limits_{n=0}^{n_\text{T}} p_\text{t}(N, n, P_{0\text{b}}, P_{1\text{b}}, \theta_\text{b}, \phi_\text{b}, \theta_\text{b} , \phi_\text{b}).
	\label{equ:pba_full_integral}
\end{align}
The acceptance probability is described by the sum of $p_\text{t}$ over all $n$ from 0 to $n_\text{T}$, as acceptance requires that the number of detected photons does not exceed
$n_\text{T}$.
To obtain the average acceptance probability, we integrate the acceptance probability $\sum \limits_{n=0}^{n_\text{T}} p_\text{t}$ over the bank angles $\theta_\text{b}, \phi_\text{b}$ using the angle probability distributions $f_{\theta}(\theta_\text{b})$, $f_{\phi}(\phi_\text{b})$.
The acceptance threshold $n_\text{T}$ can be chosen in such a way that the average probability for the bank rejecting its own generated quantum tokens is less than $\varepsilon_\text{b}$
\begin{align*}
	1-\overline{p}_\text{b} < \varepsilon_\text{b}.
\end{align*}
In the following we set $\epsilon_\text{b}=0.0002$ such that $\overline{p}_\text{b}$ of its own token is larger than $0.9998$.
We considered the test cases with $N=30$ and $N=300$ qubits in the token and performed simulations with our model and experiments on the IBMQs (see Tab.~\ref{tab:nT_threshold}).
As can be seen in Tab.~\ref{tab:nT_threshold}, one has to increase the experimental threshold $n_\text{T}^{(e)}$ compared to the simulated $n_\text{T}$ in order to compensate experimental gate errors that are not considered in our model.
In conclusion, we can see that even for the Kyoto platform with lowest contrast a high $\overline{p}_\text{b}$ can be achieved by increasing $n_\text{T}$.

\begin{table}[tb!]
	\begin{tabular}{|l||r|l||r|r|l|}
		\hline
		IBMQ  & \multicolumn{2}{c||}{$N=30$} & \multicolumn{3}{c|}{$N=300$} \\
		  &  $n_\text{T}$ & $\overline{p}_\text{b}$ & $n_\text{T}$ & $n_\text{T}^{(e)}$ & $\overline{p}_\text{b}$ \\ \hline
		Sherb. &   0 & $\sim$1 &  0 & 11 & $\sim$1 \\ 
		Kyiv   &   0 & 0.99999999998 & 0 & 10 & 0.9999999998 \\
		Osaka  &   5 & 0.99981 &  20 & 19 & 0.99982 \\ 
		Brisb. &  10 & 0.99991 &  50 & 73 & 0.99986 \\ 
		Kyoto  &  14 & 0.99986 &  83 & 96 & 0.99987 \\ 
		\hline
	\end{tabular}
	\caption{We used $N=30$ and $N=300$ qubits in the quantum token and determined the threshold $n_\text{T}$ for the different IBMQs such that the averaged bank acceptance $\overline{p}_\text{b}$ of its own token is larger than $0.9998$. The experimental threshold $n_\text{T}^{(e)}$ had to be chosen to be greater than the simulated $n_\text{T}$ due to gate errors which are not considered in our model, with only Osaka showing optimal performance ( $n_\text{T} \simeq n_\text{T}^{(e)}$).}
	\label{tab:nT_threshold}
\end{table}


\section{Attack scenarios}
\label{sec:attack}
In the following section we describe different attack scenarios together with a description of the methods used and visualizations of the attack on the Bloch-sphere. Our model assumes that a forger tries to copy the quantum token of the bank prepared with the angles $\theta_\text{b},\phi_\text{b}$, which are only known by the bank. The forger has the goal to obtain the highest possible acceptance rate by the bank for the forged token.
Forger and bank may have a different measurement setup, so that we denote the parameters of the forger setup by $P_{0\text{f}}, P_{1\text{f}}$.
In this work, we just assume that the bank has the same setup than the forger, i.e., $P_{0\text{b}}=P_{0\text{f}}$ and $P_{1\text{b}}=P_{1\text{f}}$.

\subsection{General Description of Fake Token Generation}

In this section, we provide general statements about the average acceptance probability and the averaged normalized photon count of a forged token that is generated using just a random guess or results from one, two or three measurements on the original token.
Finally, we analyze the extent to which the forger is free to choose the coordinate system and visualize the general attack scenarios on the Bloch sphere.

\subsubsection{Random Guess}

We begin with the simplest scenario: 
The forger does not perform any measurement and just guesses the angles $\theta_\text{f},\phi_\text{f}$ for preparing a forged token.
We obtain for the average probability that the bank accepts this token as a function of $\theta_\text{f},\phi_\text{f}$:
\begin{align}
	\overline{p}_{\text{f}_0} (\theta_\text{f}, \phi_\text{f}) =& \int \limits_{0}^{\pi} d\theta_\text{b} \int \limits_{-\pi}^{\pi}  d\phi_\text{b} \,  f_{\theta}(\theta_\text{b}) \, f_{\phi}(\phi_\text{b}) \, \times \nonumber \\
	 &\times  \sum \limits_{n=0}^{n_\text{T}} p_\text{t}(N, n, P_{0\text{b}}, P_{1\text{b}}, \theta_\text{f}, \phi_\text{f}, \theta_\text{b} , \phi_\text{b}).
	\label{equ:pf_0}
\end{align}
Note the similarity to Eq.~\eqref{equ:pba_full_integral}, except that in $p_\text{t}$, the fifth and sixth arguments are replaced by the angles $\theta_\text{f}$ and $\phi_\text{f}$.
The function is denoted by $\overline{p}_{\text{f}_0}$, since the forger performed 0 measurements.
Due to the uniform distribution of the bank angles over the Bloch sphere, the above probability does not depend on $\theta_\text{f},\phi_\text{f}$. 
This acceptance probability depends only on the probability $P_{0\text{b}}$ and $P_{1\text{b}}$ to measure a photon and the acceptance threshold $n_\text{T}$ of the bank (Tab.~\ref{tab:fit_params}).

For this approach, the numerical results are obtained through integration with Gauss–Legendre quadrature method \cite{Schwarz2011} and the results are presented in Sec.~\ref{sec:token_security}.

\subsubsection{One Measurement}

Now we consider the case where the forger performs one measurement on the bank's token using the angles $\theta_{\text{f}_1}, \phi_{\text{f}_1}$ and detects $n_{\text{f}_1}$ photons.
Using the information from the measurement the forger prepares a token with the angles $\theta_\text{f}$, $\phi_\text{f}$.
The average probability that the bank accepts this token is given by
\begin{align}
	\overline{p}_{\text{f}_1} =& \sum_{n_{\text{f}_1}=0}^{N_1} \int \limits_{0}^{\pi} d\theta_\text{b} \int \limits_{-\pi}^{\pi}  d\phi_\text{b} \, f_\theta(\theta_\text{b}) \, f_\phi(\phi_\text{b}) \times \nonumber \\
	& \times p_\text{t}(N_1, n_{\text{f}_1}, P_{0\text{f}}, P_{1\text{f}} , \theta_{\text{f}_1}, \phi_{\text{f}_1}, \theta_\text{b} , \phi_\text{b} ) \times \nonumber \\
	& \times \sum \limits_{n=0}^{n_\text{T}} p_\text{t}(N, n, P_{0\text{b}}, P_{1\text{b}}, \theta_\text{f}, \phi_\text{f}, \theta_\text{b} , \phi_\text{b} ).
	\label{equ:pf_1}
\end{align}
Now we have modified Eq.~\eqref{equ:pf_0} by summing the average acceptance probability over all possible single measurement outcomes $n_{\text{f}_1}$, each weighted by the probability of observing that outcome, represented by $p_\text{t}$ in the second line.
Note, that here $N_1=N$ and $\theta_\text{f}$, $\phi_\text{f}$ are functions of $N_1$, $n_{\text{f}_1}$, $\theta_{\text{f}_1}$, and $\phi_{\text{f}_1}$.
The average normalized photon count $\overline{n}$ measured by the bank as a function of the bank angle $\theta_\text{b}$ is given by
\begin{align}
	\overline{n}(\theta_\text{b}) =& \frac{1}{N}  \sum_{n_{\text{f}_1}=0}^{N_1} \int \limits_{-\pi}^{\pi}  d\phi_\text{b} \, f_\phi(\phi_\text{b}) \times \nonumber \\
	& \times p_\text{t}(N_1, n_{\text{f}_1}, P_{0\text{f}}, P_{1\text{f}} , \theta_{\text{f}_1}, \phi_{\text{f}_1}, \theta_\text{b} , \phi_\text{b} ) \times \nonumber \\
	& \times \sum \limits_{n=0}^N n\, p_\text{t}(N, n, P_{0\text{b}}, P_{1\text{b}}, \theta_\text{f}, \phi_\text{f}, \theta_\text{b} , \phi_\text{b} ).
	\label{equ:n_1}
\end{align}
The average photon count is calculated as the sum over $n = 0$ to the maximum photon number $N$, where each photon count $n$ is weighted by its corresponding probability. This probability is obtained by summing over all possible measurement outcomes $n_{\text{f}_1}$; each term is the product of the probability $p_\text{t}$ that the forger measures $n_{\text{f}_1}$ photons before forging the token (second line) and the probability that the bank measures $n$ photons from the forged token based on $n_{\text{f}_1}$ ($p_\text{t}$ in the third line). As the average photon count is only a function of $\theta_\text{b}$, we further average over $\phi_\text{b}$ by integrating with the weight function $f_\phi$. In order to normalize the average photon count, we divide by $N$.

\subsubsection{Two Measurements}

Now the forger divides the token into several parts and measures each part containing $N_j$ qubits with individual angles $\theta_{\text{f}_j}, \phi_{\text{f}_j}$ and detects $n_{\text{f}_j}$ photons.
Using the measurement results, the forger generates a forged token.
The average probability that the bank accepts a forged token prepared by the forger from two measurements is given by
\begin{align*}
	\overline{p}_{\text{f}_2} =& \sum_{n_{\text{f}_1}=0}^{N_1} \sum_{n_{\text{f}_2}=0}^{N_2} \int \limits_{0}^{\pi} d\theta_\text{b} \int \limits_{-\pi}^{\pi}  d\phi_\text{b} \, f_\theta(\theta_\text{b}) \, f_\phi(\phi_\text{b}) \times \nonumber \\
	& \times p_\text{t}(N_1, n_{\text{f}_1}, P_{0\text{f}}, P_{1\text{f}}, \theta_{\text{f}_1}, \phi_{\text{f}_1}, \theta_\text{b} , \phi_\text{b} ) \times \nonumber \\
	& \times p_\text{t}(N_2, n_{\text{f}_2}, P_{0\text{f}}, P_{1\text{f}}, \theta_{\text{f}_2}, \phi_{\text{f}_2}, \theta_\text{b} , \phi_\text{b} ) \times \nonumber \\
	& \times \sum \limits_{n=0}^{n_\text{T}} p_\text{t}(N, n, P_{0\text{b}}, P_{1\text{b}}, \theta_\text{f}, \phi_\text{f}, \theta_\text{b} , \phi_\text{b} ).
\end{align*}
In order to account for two measurements Eq.~\eqref{equ:pf_1} is modified by now summing over all two possible outcomes $n_{\text{f}_1}$ and $n_{\text{f}_2}$, and accounting for the probabilities of both outcomes, as represented by $p_\text{t}$ in lines two and three.
Here, $\theta_\text{f}, \phi_\text{f}$ are functions of $N_1$, $N_2$, $n_{\text{f}_1}$, $n_{\text{f}_2}$,  $\theta_{\text{f}_1}$, $\phi_{\text{f}_1}$, $\theta_{\text{f}_2}$, and $\phi_{\text{f}_2}$.
Now, the average normalized photon count measured by the bank as a function of the bank angle $\theta_\text{b}$ is given by
\begin{align*}
	\overline{n}(\theta_\text{b}) =& \frac{1}{N}  \sum_{n_{\text{f}_1}=0}^{N_1} \sum_{n_{\text{f}_2}=0}^{N_2} \int \limits_{-\pi}^{\pi}  d\phi_\text{b} \, f_\phi(\phi_\text{b}) \times \nonumber \\
	& \times p_\text{t}(N_1, n_{\text{f}_1}, P_{0\text{f}}, P_{1\text{f}}, \theta_{\text{f}_1}, \phi_{\text{f}_1}, \theta_\text{b} , \phi_\text{b} ) \times \nonumber \\
	& \times p_\text{t}(N_2, n_{\text{f}_2}, P_{0\text{f}}, P_{1\text{f}}, \theta_{\text{f}_2}, \phi_{\text{f}_2}, \theta_\text{b} , \phi_\text{b} ) \times \nonumber \\
	& \times \sum \limits_{n=0}^{N} n\, p_\text{t}(N, n, P_{0\text{b}}, P_{1\text{b}}, \theta_\text{f}, \phi_\text{f}, \theta_\text{b} , \phi_\text{b} ).
\end{align*}
Again two measurements are implemented through changing Eq.~\eqref{equ:n_1} by summing over all two possible outcomes $n_{\text{f}_1}$ and $n_{\text{f}_2}$, and accounting for the probabilities of both outcomes, as given in lines two and three.

\subsubsection{Three Measurements}

Finally if the forger performs three measurements, the average probability for the bank accepting the forged token is given by
\begin{align*}
	\overline{p}_{\text{f}_3} =& \sum_{n_{\text{f}_1}=0}^{N_1} \sum_{n_{\text{f}_2}=0}^{N_2} \sum_{n_{\text{f}_3}=0}^{N_3} \int \limits_{0}^{\pi} d\theta_\text{b} \int \limits_{-\pi}^{\pi}  d\phi_\text{b} \, f_\theta(\theta_\text{b}) \, f_\phi(\phi_\text{b}) \times \nonumber \\
	& \times p_\text{t}(N_1, n_{\text{f}_1}, P_{0\text{f}}, P_{1\text{f}}, \theta_{\text{f}_1}, \phi_{\text{f}_1}, \theta_\text{b} , \phi_\text{b} ) \times \nonumber \\
	& \times p_\text{t}(N_2, n_{\text{f}_2}, P_{0\text{f}}, P_{1\text{f}}, \theta_{\text{f}_2}, \phi_{\text{f}_2}, \theta_\text{b} , \phi_\text{b} ) \times \nonumber \\
	& \times p_\text{t}(N_3, n_{\text{f}_3}, P_{0\text{f}}, P_{1\text{f}}, \theta_{\text{f}_3}, \phi_{\text{f}_3}, \theta_\text{b} , \phi_\text{b} ) \times \nonumber \\
	& \times \sum \limits_{n=0}^{n_\text{T}} p_\text{t}(N, n, P_{0\text{b}}, P_{1\text{b}}, \theta_\text{f}, \phi_\text{f}, \theta_\text{b} , \phi_\text{b} ).
\end{align*}
By now the extension pattern of equation Eq. \eqref{equ:pf_1} should be obvious, we now sum over all three possible outcomes $n_{\text{f}_1}$, $n_{\text{f}_2}$, and $n_{\text{f}_3}$, and account for the probabilities of the three outcomes, as represented by $p_\text{t}$ in lines two, three, and four.
Again, $\theta_\text{f}, \phi_\text{f}$ are functions of $N_1$, $N_2$, $N_3$, $n_{\text{f}_1}$, $n_{\text{f}_2}$, $n_{\text{f}_3}$, $\theta_{\text{f}_1}$, $\phi_{\text{f}_1}$, $\theta_{\text{f}_2}$, $\phi_{\text{f}_2}$, $\theta_{\text{f}_3}$, and $\phi_{\text{f}_3}$. The choice of the angles $\theta_{\text{f}_j}, \phi_{\text{f}_j}$ and the number of qubits $N_j$ of the actual measurement may depend on the measurement results of the previous measurements.
The average normalized photon count measured by the bank as a function of the bank angle $\theta_\text{b}$ is denoted by
\begin{align*}
	\overline{n}(\theta_\text{b}) =& \frac{1}{N} \sum_{n_{\text{f}_1}=0}^{N_1} \sum_{n_{\text{f}_2}=0}^{N_2} \sum_{n_{\text{f}_3}=0}^{N_3} \int \limits_{-\pi}^{\pi}  d\phi_\text{b} \, f_\phi(\phi_\text{b}) \times \nonumber \\
	& \times p_\text{t}(N_1, n_{\text{f}_1}, P_{0\text{f}}, P_{1\text{f}}, \theta_{\text{f}_1}, \phi_{\text{f}_1}, \theta_\text{b} , \phi_\text{b} ) \times \nonumber \\
	& \times p_\text{t}(N_2, n_{\text{f}_2}, P_{0\text{f}}, P_{1\text{f}}, \theta_{\text{f}_2}, \phi_{\text{f}_2}, \theta_\text{b} , \phi_\text{b} ) \times \nonumber \\
	& \times p_\text{t}(N_3, n_{\text{f}_3}, P_{0\text{f}}, P_{1\text{f}}, \theta_{\text{f}_3}, \phi_{\text{f}_3}, \theta_\text{b} , \phi_\text{b} ) \times \nonumber \\
	& \times \sum \limits_{n=0}^{N} n\, p_\text{t}(N, n, P_{0\text{b}}, P_{1\text{b}}, \theta_\text{f}, \phi_\text{f}, \theta_\text{b} , \phi_\text{b} ).
\end{align*}
And as previously we have extended Eq.~\eqref{equ:n_1} by summing over all three possible outcomes $n_{\text{f}_1}$, $n_{\text{f}_2}$, and $n_{\text{f}_3}$, and accounting for the probabilities of the three outcomes, as represented by $p_\text{t}$ in lines two, three, and four.

\subsubsection{Forger's Measurement Process}

\begin{figure}[htb!]
    \centering
    \includegraphics[width=0.9\columnwidth]{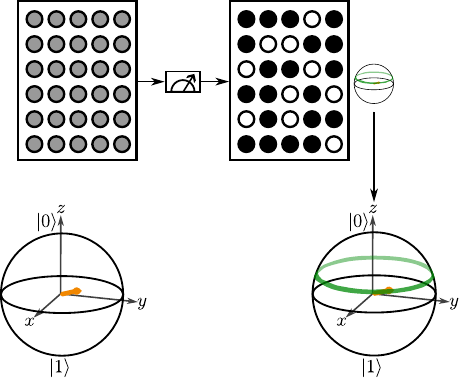}
    \caption{The forger performs a single measurement on the entire quantum token, resulting in some qubits emitting a photon, as indicated by the white filled circles. On the Bloch sphere, the states yielding this measurement outcome with highest probability are distributed along a ring.
    }
    \label{fig:measurement_1}
\end{figure}

\begin{figure*}[htb!]
    \centering
    \includegraphics[width=0.7\textwidth]{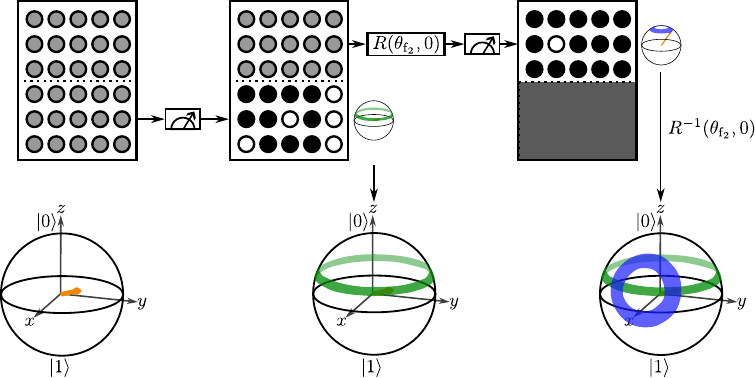}
    \caption{An improved measurement scheme consists of two measurements. Now the forger divides the token into two parts. First, one part is measured as in Fig.~\ref{fig:measurement_1}. Note that the width of the ring is now larger as compared to a single measurement due to the reduced ensemble size.  The forger then measures the second part of the token after applying a rotation by $\theta_{\text{f}_2}$, leading now to a typically different positioned ring (see top blue ring). By applying the inverse rotation $-\theta_{\text{f}_2}$, this second result can be visualized together with the first measurement. The intersection of the two rings generally consists of two distinct regions, which represent the states on the Bloch sphere most likely to produce both measurement outcomes. Even though the area of possible angles was reduced compared to a single measurement, there still exists an ambiguity.
    }
    \label{fig:measurement_2}
\end{figure*}

\begin{figure*}[t!]
    \centering
    \includegraphics[width=0.98\textwidth]{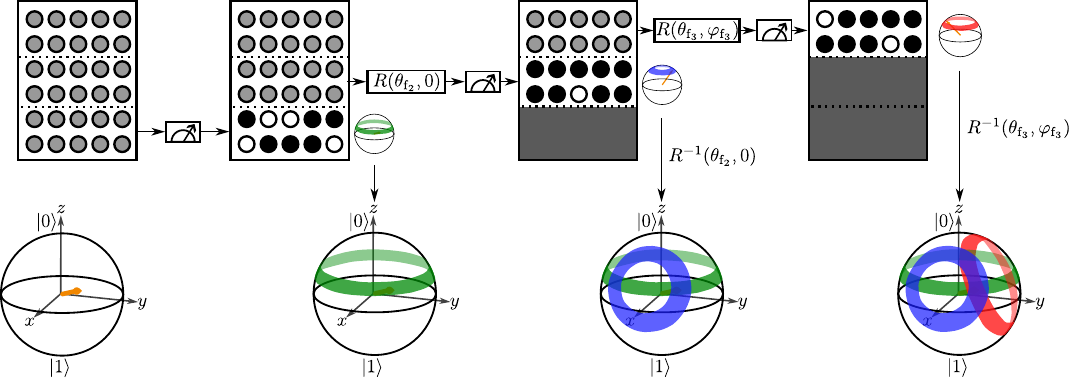}
    \caption{With three measurements typically an unambiguous result is obtained. Now the forger divides the token into three parts. The forger measures each part but he rotates the second and third sub-ensembles by angles $\theta_{\text{f}_2}$ and $\theta_{\text{f}_3}$, $\phi_{\text{f}_3}$ respectively. Note that the width of the ring is now larger as compared to a single and two measurements due to the reduced ensemble size. But after undoing the rotation by $\theta_{\text{f}_3}$ and $\phi_{\text{f}_3}$, typically, only a single distinct intersection region exists, which contains the states most likely to produce all three measurement outcomes. With this method typically an unambiguous result with minimal area of possible angles is obtained.
    }
    \label{fig:measurement_3}
\end{figure*}

Since the bank chooses the angles $\theta_\text{b},\phi_\text{b}$ uniformly distributed on the Bloch sphere, as described in Eq.~\ref{equ:fbtheta_fbphi}, there are no preferred directions and the forger has the free choice of setting the measurement basis.
Thus, the forger chooses the basis in such a way that the angles $\theta_{\text{f}_1}=0$, $\phi_{\text{f}_1}=0$ are used  in the first measurement and the angle $\phi_{\text{f}_2}=0$ with arbitrary $\theta_{\text{f}_2}$ is used  in the second measurement.
In the last case $\phi_{\text{f}_2}=0$ is no restriction of generality because the value of $\phi_{\text{f}_1}$ does not play any role when measured at the pole with $\theta_{\text{f}_1}=0$ (see Eq.~\eqref{equ:p_qubit_fin}). 
In Fig.~\ref{fig:measurement_1}, Fig.~\ref{fig:measurement_2}, and Fig.~\ref{fig:measurement_3}, we visualize the measurement process for one, two and three measurements.

\subsection{Quantum State Tomography}

In the general description of fake token generation, the forger has to concretely choose the measurement angles of the second and third measurement and has to define the preparation angles of the forged token $\theta_\text{f}$, $\phi_\text{f}$ from the measurement results.
In this section, we describe several strategies determining these angles depending on the measurement result using state-of-the art quantum state tomography methods. Here, the forger tries to determine $\theta_\text{f}$, $\phi_\text{f}$ from the measurements trying to reach values as close as possible to the unknown bank angles $\theta_\text{b}$, $\phi_\text{b}$.
We compare three different methods for quantum state tomography: direct inversion tomography, maximum likelihood method and Bayesian method.


\subsubsection{Direct Inversion Tomography (DIT)}

\begin{figure}[htb!]
    \centering
    \includegraphics[width=\columnwidth]{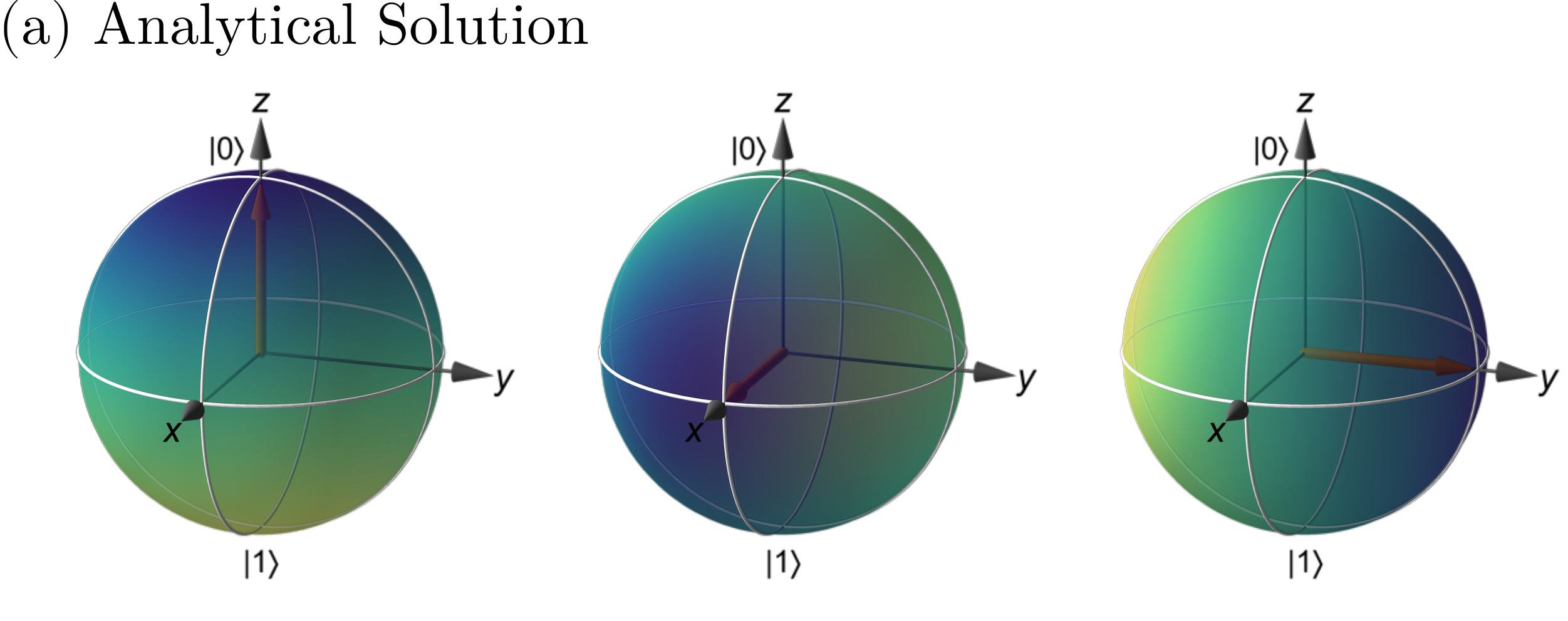}
    \includegraphics[width=\columnwidth]{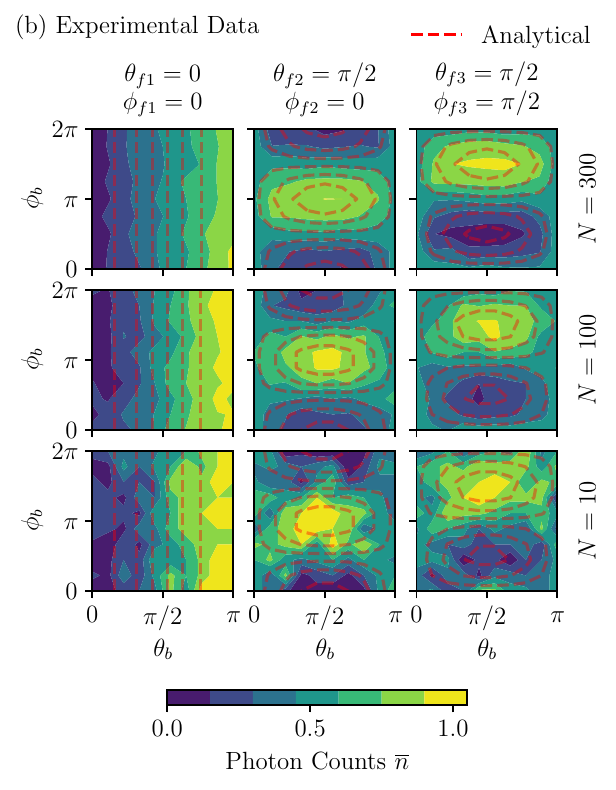}
    \caption{(a) Analytic solution from Eq.~\ref{equ:p_qubit_fin} for the averaged normalized photon counts $\overline{n}$ which an attacker obtains by measuring a bank token along the $z$, $x$ and $y$ axes. As the bank's angles approach the attacker's angles in the Bloch sphere, the photon counts get smaller. (b) Measurements performed by the attacker on IBMQ Brisbane with $N=300$, 100 and 10 qubits along the three axes on the Bloch sphere and projected on a 2D-plane. The data is compared with the analytic solution for the fitted value of $P_0$ and $P_1$ obtained from the Rabi measurement in Fig.~\ref{fig:fit_params}. As the ensemble size approaches a small value of $N=10$, the measurement becomes noisy and deviates from theory. Whereas at $N=100$ and 300, a good agreement is observed.}
    \label{fig:tomography}
\end{figure}

The simplest method providing in principle the complete information of the bank state $\theta_\text{b},\phi_\text{b}$ is performing three measurements, one in each of the dimensions on the Bloch sphere using $N_j=N/3$ qubits.
In detail, the forger measures $n_{\text{f}_1}$ photons using $\theta_{\text{f}_1}=0$, $\phi_{\text{f}_1}=0$, $n_{\text{f}_2}$ photons using $\theta_{\text{f}_2}=\frac{\pi}{2}$, $\phi_{\text{f}_2}=0$ and $n_{\text{f}_3}$ photons using $\theta_{\text{f}_3}=\frac{\pi}{2}$, $\phi_{\text{f}_3}=\frac{\pi}{2}$.
This corresponds to a measurement along the $z$-, $x$- and $y$-axis, as shown in Fig.~\ref{fig:tomography} (a).

Using Eq.~\ref{equ:p_qubit_fin} for the ideal case where $P_0=0$ and $P_1=1$, 
$\overline{n}$ can be calculated for this measurement scheme as a function of the bank angles $\theta_\text{b}$ and $\phi_\text{b}$.
These results plotted over the Bloch sphere are shown in Fig.~\ref{fig:tomography} (a). 
We observe that the number of detected counts decreases as the measurement angles chosen by the attacker approach the secret angles used by the bank. To further test the model, this measurement procedure was performed with IBMQ Brisbane for three ensemble sizes of $N=300$, 100 and 10, as shown in Fig.~\ref{fig:tomography} (b). The Bloch sphere is projected into a 2D-plane for better visualization, where we observe a good agreement with the experimental data for the largest ensemble size. As the ensemble size decreases to $N=10$, the measurement becomes increasingly noisy. Therefore, we chose an intermediate ensemble size of $N=300$ allowing for low noise state estimation if we take 100 measurements in each axis. Additionally, we realized that the photon counts are independent of $\phi_\text{b}$ for the measurement in the $z$-axis, as expected. In the following, the forger uses this measurement results to forge the fake tokens.

From the three results $n_{\text{f}_1}$, $n_{\text{f}_2}$, $n_{\text{f}_3}$, the forger obtains the following guess of the bank's state Bloch vector \cite{Schmied2016}
\begin{align*}
    \mathbf{R}_\text{d} =& \left(\frac{ N_2 - 2\,n_{\text{f}_2} }{N_2}, \frac{ N_3 - 2\,n_{\text{f}_3}}{N_3}, \frac{ N_1 - 2\,n_{\text{f}_1}}{N_1} \right).
\end{align*}

In general, $||\mathbf{R}_\text{d}|| \not= 1$, so that the forger has to perform a normalization in order to obtain a physically reasonable guess of the bank's state
\begin{align*}
    \mathbf{r}_\text{d} = \frac{\mathbf{R}_\text{d}}{||\mathbf{R}_\text{d}||}.
\end{align*}
Finally, the forger obtains the angles $\theta_\text{f}^{(\text{DIT})}, \phi_\text{f}^{(\text{DIT})}$ for the best guess of the bank's angles from
\begin{align}
    \theta_\text{f}^{(\text{DIT})} = \text{arccos}\bigl( r_\text{d}^{(z)} \bigr),
    \label{equ:theta_f_DIT}
\end{align}
where arccos denotes the inverse cosine function and $r_\text{d}^{(z)}$ is the $z$-component of the vector $\mathbf{r}_\text{d}$.
If $\theta_\text{f}^{(\text{DIT})} \in \{0,\pi\}$, we can set $\phi_\text{f}^{(\text{DIT})} = 0$, since the value of $\phi_\text{f}^{(\text{DIT})}$ does not play any role.
Otherwise, we obtain
\begin{align}
    \phi_\text{f}^{(\text{DIT})} =& \text{atan2}\left( \frac{r_\text{d}^{(\text{x})}}{ \sin\bigl( \theta_\text{f}^{(\text{DIT})} \bigr) }, \frac{r_\text{d}^{(\text{y})}}{ \sin\bigl( \theta_\text{f}^{(\text{DIT})} \bigr) } \right),
    \label{equ:phi_f_DIT}
\end{align}
where $\text{atan2}$ is the 2-argument arctangent function.
Note that $\theta_\text{f}^{(\text{DIT})}, \phi_\text{f}^{(\text{DIT})}$ are functions of $n_{\text{f}_1}$, $n_{\text{f}_2}$ and $n_{\text{f}_3}$.
We denote the average probability for the bank accepting this forged token by $\overline{p}_{\text{f}_3}^{(\text{DIT})}$ and present our numerical results in Sec.~\ref{sec:token_security}.

\begin{figure}[htb!]
    \centering
    \includegraphics[width=\columnwidth]{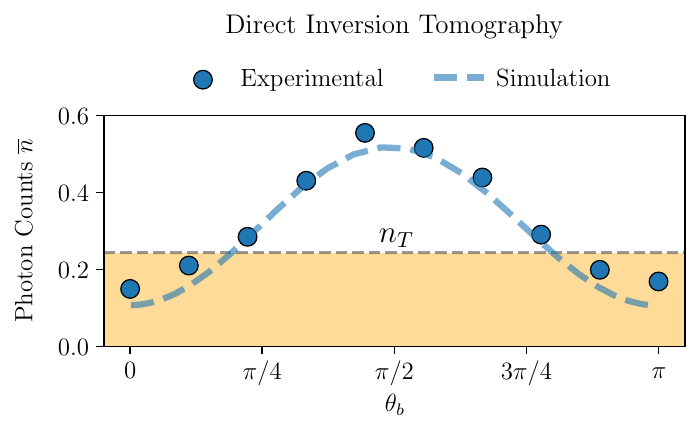}
    \caption{IBMQ Brisbane simulated and experimental normalized photon counts measured by the bank from tokens forged with direct inversion tomography. Using the measurement results from Fig.~\ref{fig:tomography}, the attacker sets the forged angles $\theta_\text{f}^{(\text{DITe})}$ and $\phi_\text{f}^{(\text{DITe})}$ for the token, which is then passed to the bank. The photon counts measured by the bank and averaged over $\phi_\text{b}$ show a maximum at $\theta_\text{b}=\pi/2$, due to the larger distribution of states at the equator of the Bloch sphere. The token is accepted if the photon counts are smaller than the threshold $\overline{n} < n_\text{T}$, where $n_\text{T}^\text{(e)}$ taken from Tab.~\ref{tab:nT_threshold}. Averaging over $\theta_\text{b}$ according to the distribution of Eq.~\eqref{equ:fbtheta_fbphi}, we obtain an acceptance probability of $\overline{p}_{\text{f}_3}^{(\text{DITe})} = 0.3769$ for tokens forged by DIT.}
    \label{fig:nb_DIT}
\end{figure}

Forged tokens were experimentally prepared with this method based on the attacker measurements from Fig.~\ref{fig:tomography}~(b) for $N=300$, with 100 qubits measured in each axis. Subsequently these forged tokens are passed to the bank, where the photons counts are measured with the original angles $\theta_\text{b}, \phi_\text{b}$.
These normalized photon counts obtained with IBMQ Brisbane as a function of the preparation angle $\theta_\text{b}$ and averaged over $\phi_\text{b}$ are shown in Fig.~\ref{fig:nb_DIT}. 

As discussed in Sec.~\ref{sec:bank}, the token is accepted if the photon counts are smaller than the threshold $\overline{n} < n_\text{T}^\text{(e)}$, where $n_\text{T}^\text{(e)}$ is taken from Tab.~\ref{tab:nT_threshold}. Due to smaller density of states at the poles we observe that the acceptance probability is larger, in agreement with Eq.~\ref{equ:fbtheta_fbphi}. The experimental data agrees well with the numerical values. Averaging over $\theta_b$ weighted according to the spherical distribution (Eq.~\ref{equ:fbtheta_fbphi}), we obtain the experimental acceptance probability of $\overline{p}_{\text{f}_3}^{(\text{DITe})} = 0.3769$, which is higher than the numerical value $\overline{p}_{\text{f}_3}^{(\text{DIT})} = 0.2770$, due to the different acceptance thresholds $n_\text{T}^{(\text{e})} > n_\text{T} $. We will present in Sec.~\ref{sec:token_security} a detailed analysis of the numerical results.


\subsubsection{Maximum Likelihood Method (ML)}

The above method has the disadvantage that the forger has to divide the quantum token into three parts and has to perform a measurement with different angles on each part. 
This may be technically unfeasible for some quantum systems like NV-centers, where the ensemble inseparability is guaranteed by the diffraction limited area of the optical initialization and readout \cite{optical_pumping}, as well as the non-local microwave state manipulation \cite{rabi_NV}.
Another approach is using the maximum likelihood method, which can also be used with only one measurement on the entire token.
In this method, the forger tries to determine $\theta_\text{f}$, $\phi_\text{f}$ in such a way that these angles most likely generate the observed measurement results.
The forger performs $N_\text{m}$ measurements and obtains $n_{\text{f}_j}$ photons in the $j$-th measurement with the angles $\theta_{\text{f}_j}$, $\phi_{\text{f}_j}$ on $N_j$ qubits.
The likelihood function, which describes the probability of the total experimental outcomes, is given by \cite{Paris2016}
\begin{align}
    \mathcal{L}\bigl( \{ n_{\text{f}_j} \} \bigl| \theta_\text{b},\phi_\text{b} \bigr) =& \prod \limits_{j=1}^{N_\text{m}} p_\text{t}(N_j,n_{\text{f}_j}, P_{0\text{f}},P_{1\text{f}}, \theta_{\text{f}_j},\phi_{\text{f}_j}, \theta_\text{b}, \phi_\text{b}).
    \label{equ:Likelihood}
\end{align}
The forger searches for the angles $\theta_\text{b}$, $\phi_\text{b}$ that maximize the likelihood function, which can be found through the conditions
\begin{align*}
    \frac{\partial \mathcal{L}}{\partial \theta_\text{b}} =0, \qquad \frac{\partial \mathcal{L}}{\partial \phi_\text{b}} = 0.
\end{align*}
The solution provides the maximum likelihood estimation $\theta_\text{f}^{(\text{ML})}$, $\phi_\text{f}^{(\text{ML})}$ for the unknown bank angles $\theta_\text{b}$, $\phi_\text{b}$.
Thus, in order to determine the solution, we formally exchange $\theta_\text{b}$, $\phi_\text{b}$ with  $\theta_\text{f}^{(\text{ML})}$, $\phi_\text{f}^{(\text{ML})}$ in Eq.~\eqref{equ:Likelihood}.

Firstly, we consider the case of $N_\text{m}=1$ measurement.
Here, we have 
\begin{align*}
    & \mathcal{L}\bigl( n_{\text{f}_1} \bigl| \theta_\text{f}^{(\text{ML})}, \phi_\text{f}^{(\text{ML})} \bigr) \nonumber \\
    =& p_\text{t}\bigl(N_1,n_{\text{f}_1}, P_{0\text{f}},P_{1\text{f}}, 0,0,  \theta_\text{f}^{(\text{ML})}, \phi_\text{f}^{(\text{ML})}\bigr)
\end{align*}
using $\theta_{\text{f}_1}=0$, $\phi_{\text{f}_1}=0$, and $N_1=N$. Since $p_\text{t}$ is a binomial distribution with probability $p_\text{q}\bigl(P_{0\text{f}},P_{1\text{f}},0,0, \theta_\text{f}^{(\text{ML})}, \phi_\text{f}^{(\text{ML})}\bigr)$, the likelihood function is maximal for \cite{Handl2018}
\begin{align*}
    \frac{n_{\text{f}_1}}{N} =& p_\text{q}\bigl(P_{0\text{f}},P_{1\text{f}}, 0, 0, \theta_\text{f}^{(\text{ML})}, \phi_\text{f}^{(\text{ML})}\bigr).
\end{align*}
Using Eq.~\eqref{equ:p_qubit_fin}, one obtains further
\begin{align*}
	\frac{n_{\text{f}_1}}{N_1}=&\frac{P_0+P_1}{2} - \frac{P_1-P_0}{2} \cos\bigl(\theta_\text{f}^{(\text{ML})}\bigr).
\end{align*}
A rearrangement yields
\begin{align}
	\underbrace{\frac{2\,\frac{n_{\text{f}_1}}{N_1} - P_0 - P_1}{P_0-P_1}}_{=:a_{\text{f}_1}} = \cos\bigl(\theta_\text{f}^{(\text{ML})}\bigr),
	\label{equ:af1}
\end{align}
where we get
\begin{align}
	\theta_\text{f}^{(\text{ML})} = \arccos( a_{\text{f}_1} ).
	\label{equ:ML_1m_theta_ML}
\end{align}
Due to noise in a real world measurement, one may obtain a value of $a_{\text{f}_1}$ that is outside of the interval $[-1,1]$.
In order to obtain a physically meaningful value for $\theta_\text{f}^{(\text{ML})}$ in these cases, $a_{\text{f}_1}$ must be constrained to lie within the interval $[-1,1]$.
For $\phi_\text{f}^{(\text{ML})}$, the forger can choose any value in the interval $(-\pi,\pi]$. We choose the value $\phi_\text{f}^{(\text{ML})}=0$ without restriction of generality.

Now we consider $N_\text{m}=2$ measurements with $\theta_{\text{f}_2} \notin \{0,\pi\}$.
Note, for $\theta_{\text{f}_2} \in \{0,\pi\}$, the second measurement is equivalent to the first measurement, so that one effectively performs one measurement on the whole quantum token as described above.
We have
\begin{align*}
    &\mathcal{L}\bigl( n_{\text{f}_1}, n_{\text{f}_2} \bigl| \theta_\text{f}^{(\text{ML})}, \phi_\text{f}^{(\text{ML})} \bigr) \nonumber \\
    =& p_\text{t}\bigl(N_1,n_{\text{f}_1}, P_{0\text{f}},P_{1\text{f}}, 0,0, \theta_\text{f}^{(\text{ML})}, \phi_\text{f}^{(\text{ML})} \bigr) \times \nonumber \\
    & \times p_\text{t}\bigl(N_2,n_{\text{f}_2}, P_{0\text{f}},P_{1\text{f}}, \theta_{\text{f}_2},0, \theta_\text{f}^{(\text{ML})}, \phi_\text{f}^{(\text{ML})} \bigr), 
\end{align*}
Using $\theta_{\text{f}_1}=0$, $\phi_{\text{f}_1}=0$, and $\phi_{\text{f}_2}=0$.
The likelihood function is maximal, if the two factors are maximal. 
This generates directly the conditions
\begin{align}
    \frac{n_{\text{f}_1}}{N_1} =& p_\text{q}\bigl(P_{0\text{f}},P_{1\text{f}}, 0, 0, \theta_\text{f}^{(\text{ML})}, \phi_\text{f}^{(\text{ML})} \bigr), 
    \label{equ:ML_2m_cond1}\\
    \frac{n_{\text{f}_2}}{N_2} =& p_\text{q}\bigl(P_{0\text{f}},P_{1\text{f}}, \theta_{\text{f}_2}, 0, \theta_\text{f}^{(\text{ML})}, \phi_\text{f}^{(\text{ML})}\bigr).
    \label{equ:ML_2m_cond2}
\end{align}
From Eq.~\eqref{equ:ML_2m_cond1}, we can directly derive $\theta_\text{f}^{(\text{ML})}$ via Eq.~\eqref{equ:ML_1m_theta_ML}.
Using $\theta_\text{f}^{(\text{ML})}$, we obtain from Eq.~\eqref{equ:ML_2m_cond2}
\begin{align}
    a_{\text{f}_2} =& \cos(\theta_{\text{f}_2}) \, \cos\bigl(\theta_\text{f}^{(\text{ML})}\bigr) \nonumber \\
    & + \sin(\theta_{\text{f}_2}) \, \sin\bigl(\theta_\text{f}^{(\text{ML})}\bigr) \, \cos\bigl(\phi_\text{f}^{(\text{ML})}\bigr),
\end{align}
if we define $a_{\text{f}_2}$ as a function of $N_2$ and $n_{\text{f}_2}$ by replacing the corresponding variables in $a_{\text{f}_1}$ following Eq.~\eqref{equ:af1}.
If $\theta_\text{f}^{(\text{ML})} \in \{0, \pi\}$, then $\sin\bigl(\theta_\text{f}^{(\text{ML})}\bigr)=0$ and the value of  $\phi_\text{f}^{(\text{ML})}$ does not play any role, so that we can set $\phi_\text{f}^{(\text{ML})}=0$.
Otherwise, a rearrangement yields
\begin{align*}
    \frac{a_{\text{f}_2} - \cos(\theta_{\text{f}_2}) \, \cos\bigl(\theta_\text{f}^{(\text{ML})}\bigr)}{ \sin(\theta_{\text{f}_2}) \, \sin\bigl(\theta_\text{f}^{(\text{ML})}\bigr)} =& \cos\bigl(\phi_\text{f}^{(\text{ML})}\bigr).
\end{align*}
Since we have $\cos\bigl(\theta_\text{f}^{(\text{ML})}\bigr) = a_{\text{f}_1}$ and
\begin{align*}
    \sin\bigl(\theta_\text{f}^{(\text{ML})}\bigr) = \sqrt{1-\cos\bigl(\theta_\text{f}^{(\text{ML})}\bigr)^2} = \sqrt{1 - a_{\text{f}_1}^2},
\end{align*}
we obtain
\begin{align}
    \phi_\text{f}^{(\text{ML}\pm)} = \pm\text{arccos}\left( \frac{a_{\text{f}_2} - \cos(\theta_{\text{f}_2}) \, a_{\text{f}_1} }{ \sin(\theta_{\text{f}_2}) \, \sqrt{1 - a_{\text{f}_1}^2} } \right).
    \label{equ:ML_2m_phi_ML_pm}
\end{align}
Again, one has to adapt the argument of the arccos function so that it is located in the interval $[-1,1]$ in order to obtain real solutions.
In general, there are two solutions for $\phi_\text{f}^{(\text{ML})}$, from which the forger has to choose one. Due to symmetry properties, both solutions are equivalent, such that we can choose the "+" solution without restriction of generality.

At last, we consider $N_\text{m}=3$ measurements with  $\theta_{\text{f}_2} \notin \{0,\pi\}$ and $\theta_{\text{f}_3} \notin \{0,\pi\}$ and $\phi_{\text{f}_3} \notin \{0,\pi\}$.
The constraints $\theta_{\text{f}_2} \notin \{0,\pi\}$ and $\theta_{\text{f}_3} \notin \{0,\pi\}$ are responsible for all measurements being different.
The constraint $\phi_{\text{f}_3} \notin \{0,\pi\}$ takes care that the third measurement is not equivalent to the second measurement.
We have
\begin{align*}
    &\mathcal{L}\bigl( n_{\text{f}_1}, n_{\text{f}_2}, n_{\text{f}_3} \bigl| \theta_\text{f}^{(\text{ML})}, \phi_\text{f}^{(\text{ML})} \bigr) \nonumber \\
    =& p_\text{t}\bigl(N_1,n_{\text{f}_1}, P_{0\text{f}},P_{1\text{f}}, 0,0, \theta_\text{f}^{(\text{ML})}, \phi_\text{f}^{(\text{ML})} \bigr) \times \nonumber \\
    & \times p_\text{t}\bigl(N_2,n_{\text{f}_2}, P_{0\text{f}},P_{1\text{f}}, \theta_{\text{f}_2},0, \theta_\text{f}^{(\text{ML})}, \phi_\text{f}^{(\text{ML})} \bigr) \times \nonumber \\
    & \times p_\text{t}\bigl(N_3,n_{\text{f}_3}, P_{0\text{f}},P_{1\text{f}}, \theta_{\text{f}_3},\phi_{\text{f}_3}, \theta_\text{f}^{(\text{ML})}, \phi_\text{f}^{(\text{ML})} \bigr),
\end{align*}
using $\theta_{\text{f}_1}=0$, $\phi_{\text{f}_1}=0$, and $\phi_{\text{f}_2}=0$.
The likelihood function is maximal, if the three factors are maximal, which produces the conditions
\begin{align}
    \frac{n_{\text{f}_1}}{N_1} =& p_\text{q}\bigl(P_{0\text{f}},P_{1\text{f}}, 0, 0, \theta_\text{f}^{(\text{ML})}, \phi_\text{f}^{(\text{ML})} \bigr), 
    \label{equ:ML_3m_cond1}\\
    \frac{n_{\text{f}_2}}{N_2} =& p_\text{q}\bigl(P_{0\text{f}},P_{1\text{f}}, \theta_{\text{f}_2}, 0, \theta_\text{f}^{(\text{ML})}, \phi_\text{f}^{(\text{ML})}\bigr), 
    \label{equ:ML_3m_cond2} \\
    \frac{n_{\text{f}_3}}{N_3} =& p_\text{q}\bigl(P_{0\text{f}},P_{1\text{f}}, \theta_{\text{f}_3}, \phi_{\text{f}_3}, \theta_\text{f}^{(\text{ML})}, \phi_\text{f}^{(\text{ML})}\bigr).
    \label{equ:ML_3m_cond3}
\end{align}
From Eq.~\eqref{equ:ML_3m_cond1} we obtain $\theta_\text{f}^{(\text{ML})}$ using Eq.~\eqref{equ:ML_1m_theta_ML}. Combining this result with Eq.~\eqref{equ:ML_3m_cond2} we derive the two solutions $\phi_\text{f}^{(\text{ML}\pm)}$ for $\phi_\text{f}^{(\text{ML})}$ using Eq.~\eqref{equ:ML_2m_phi_ML_pm}.
We obtain further from Eq.~\eqref{equ:ML_3m_cond3}
\begin{align*}
    \frac{a_{\text{f}_3} - \cos(\theta_{\text{f}_3}) \, \cos\bigl(\theta_\text{f}^{(\text{ML})}\bigr)}{ \sin(\theta_{\text{f}_3}) \, \sin\bigl(\theta_\text{f}^{(\text{ML})}\bigr)} =& \cos\bigl(\phi_{\text{f}_3} - \phi_\text{f}^{(\text{ML})}\bigr).
\end{align*}
Using
\begin{align*}
    & \cos\bigl(\phi_{\text{f}_3} - \phi_\text{f}^{(\text{ML})}\bigr) \nonumber \\
    =& \cos(\phi_{\text{f}_3}) \, \cos\bigl(\phi_\text{f}^{(\text{ML})}\bigr) + \sin(\phi_{\text{f}_3}) \, \sin\bigl(\phi_\text{f}^{(\text{ML})}\bigr)
\end{align*}
we further obtain
\begin{align*}
    &\frac{a_{\text{f}_3} - \cos(\theta_{\text{f}_3}) \, \cos\bigl(\theta_\text{f}^{(\text{ML})}\bigr)}{ \sin(\theta_{\text{f}_3}) \, \sin(\phi_{\text{f}_3}) \, \sin\bigl(\theta_\text{f}^{(\text{ML})}\bigr)} - \frac{\cos(\phi_{\text{f}_3}) \, \cos\bigl(\phi_\text{f}^{(\text{ML})}\bigr)}{\sin(\phi_{\text{f}_3})} \nonumber \\
    =& \sin\bigl(\phi_\text{f}^{(\text{ML})}\bigr)
\end{align*}
and finally
\begin{align*}
    & \sin\bigl(\phi_\text{f}^{(\text{ML})}\bigr) \nonumber \\
    =&\frac{a_{\text{f}_3} - \cos(\theta_{\text{f}_3}) \, a_{\text{f}_1} }{ \sin(\theta_{\text{f}_3}) \, \sin(\phi_{\text{f}_3}) \, \sqrt{1 -a_{\text{f}_1}^2 } } - \frac{\cos(\phi_{\text{f}_3}) \, \cos\bigl(\phi_\text{f}^{(\text{ML})}\bigr)}{\sin(\phi_{\text{f}_3})}.
\end{align*}
In contrast to the cosine function, the sine function is sensitive to the sign of the argument, so that one can determine the final solution $\phi_\text{f}^{(\text{ML})}$ with the correct sign by inserting the two solutions $\phi_\text{f}^{(\text{ML}\pm)}$ in the above equation.

Similar to the direct inversion tomography, the forger obtains a clear estimate of the unknown bank angles using three measurements with the maximum likelihood method.
Using two measurements, the forger obtains two possible points on the Bloch sphere and using one measurement, the forger obtains a circle on the Bloch sphere as possible solutions for the bank angles (see Fig.~\ref{fig:measurement_1}, and Fig.~\ref{fig:measurement_2}). We denote the average probability that the bank accepts the forged tokens generated by the maximum likelihood method using one, two, or three measurements as $\overline{p}_{\text{f}_1}^{(\text{ML})}$, $\overline{p}_{\text{f}_2}^{(\text{ML})}$ and $\overline{p}_{\text{f}_3}^{(\text{ML})}$, respectively.
For Brisbane and $N=300$, we obtain $\overline{p}_{\text{f}_1}^{(\text{ML})} = 0.2835$, $\overline{p}_{\text{f}_2}^{(\text{ML})} = 0.6404$ and $\overline{p}_{\text{f}_3}^{(\text{ML})}=0.9803$.
In Sec.~\ref{sec:token_security}, we present and discuss all numerical results in detail.

\begin{figure}[t!]
    \centering
    \includegraphics[width=\columnwidth]{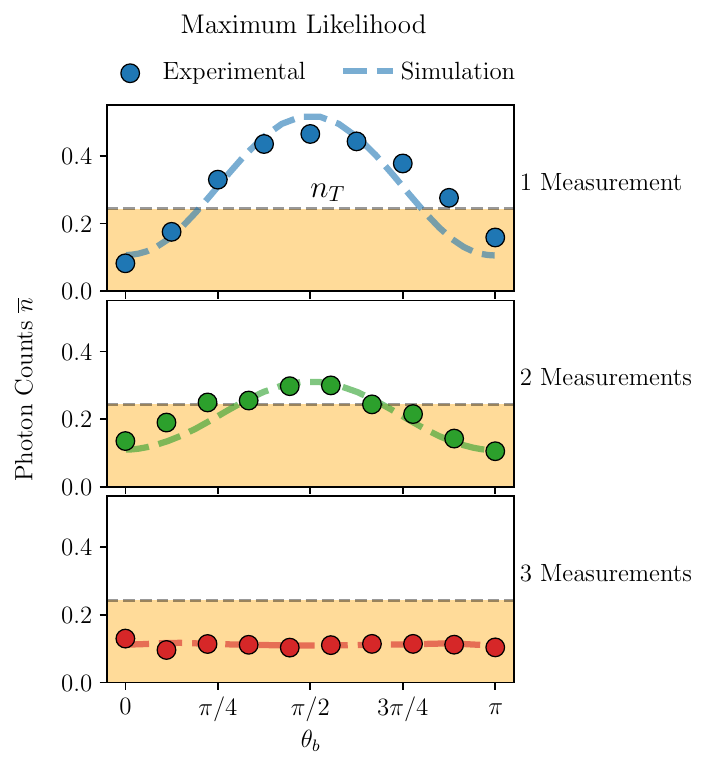}
    \caption{Simulated and experimental averaged photon counts of IBMQ Brisbane for tokens forged by maximum likelihood method for 1, 2 and 3 measurement axis performed by the forger. As in Fig.~\ref{fig:nb_DIT} for the DIT method, the forger uses the results from his measurements to forge fake tokens for many combinations of bank angles $\theta_\text{b}$ and $\phi_\text{b}$ of the token, which is then passed to the bank and verified. The photon counts are maximal at $\theta_\text{b}=\pi/2$ for 1 and 2 measurements, due to the larger density of states at this point. While for 3 measurements, the attacker can efficiently forge fake tokens over the whole Bloch sphere. The corresponding experimental acceptance probabilities are $\overline{p}_{\text{f}_1}^{(\text{MLe})}=0.4698$ $\overline{p}_{\text{f}_2}^{(\text{MLe})}=0.6688$ and $\overline{p}_{\text{f}_3}^{(\text{MLe})}=0.9847$.}
    \label{fig:nb_ML}
\end{figure}

These values were also measured experimentally with IBMQ Brisbane using $N=300$, as shown in Fig.~\ref{fig:nb_ML}. As in the case of the DIT method, the normalized counts are maximal at $\theta_\text{b}=\pi/2$. 
As the number of measurement axes are increased, the counts get lower at the equator resulting in a better estimation of the state by the forger.
By averaging over $\theta_\text{b}$, weighted by Eq.~\ref{equ:fbtheta_fbphi} and assuming the experimental benchmarked threshold of $n_\text{T}^\text{(e)}$ from Tab.~\ref{tab:nT_threshold}, we get the acceptance probabilities of $\overline{p}_{\text{f}_1}^{(\text{MLe})}=0.4698$ $\overline{p}_{\text{f}_2}^{(\text{MLe})}=0.6688$ and $\overline{p}_{\text{f}_3}^{(\text{MLe})}=0.9847$. 
From this results one can clearly see that the acceptance probability of the forged token increases substantially with the number of measurements. 
Note that several measurements might not be physically realizable on every quantum platform.

\subsubsection{Bayesian Method (Ba)}

If the forger performs multiple measurements on the token, the outcomes of previous measurements can be used to optimize subsequent measurement settings via the Bayesian update rule. 
For this the ensemble should be split into sub-ensembles.
Before the $j$-th measurement, the knowledge of the forger is described by a prior probability distribution $p^{(j-1)}(\theta_\text{b},\phi_\text{b})$ of the bank angles.
In detail, before any measurement, the forger can only assume uniform distribution of the bank angles from Eq.~\eqref{equ:fbtheta_fbphi}, i.e.,
\begin{align*}
    p^{(0)}(\theta_\text{b},\phi_\text{b}) = f_\theta(\theta_\text{b}) \, f_\phi (\phi_\text{b}).
\end{align*}
Now the forger performs the $j$-th measurement on $N_j$ qubits using the angles $\theta_{\text{f}_j}$, $\phi_{\text{f}_j}$ and measures $n_{\text{f}_j}$ photons.
From the result $n_{\text{f}_j}$, the forger can derive the posterior probability distribution of the bank angles from the Bayesian update rule \cite{Georg2016} via
\begin{align*}
    p^{(j)}(\theta_\text{b},\phi_\text{b}) 
    :=& p\bigl((\theta_\text{b}, \phi_\text{b}) \bigl| n_{\text{f}_j}, (\theta_{\text{f}_j}, \phi_{\text{f}_j}) \bigr) \nonumber \\
    =& \frac{p\bigl(n_{\text{f}_j} \bigl| (\theta_\text{b}, \phi_\text{b}), (\theta_{\text{f}_j}, \phi_{\text{f}_j}) \bigr) \, p^{(j-1)}(\theta_\text{b}, \phi_\text{b})}{p\bigl(n_{\text{f}_j} \bigl| (\theta_{\text{f}_j}, \phi_{\text{f}_j}) \bigr)},
\end{align*}
where $p(n_{\text{f}_j} | (\theta_\text{b}, \phi_\text{b}), (\theta_{\text{f}_j}, \phi_{\text{f}_j}) )$ is the conditional probability of observing $n_{\text{f}_j}$ photons, if one measures in the basis  $\theta_{\text{f}_j}, \phi_{\text{f}_j}$ a token that is prepared with angles $\theta_\text{b}, \phi_\text{b}$, which is given by
\begin{align*}
    & p\bigl(n_{\text{f}_j} \bigl| (\theta_\text{b}, \phi_\text{b}), (\theta_{\text{f}_j}, \phi_{\text{f}_j}) \bigr) \nonumber \\
    =& p_\text{t}(N_j, n_{\text{f}_j}, P_{0\text{f}}, P_{1\text{f}}, \theta_{\text{f}_j}, \phi_{\text{f}_j}, \theta_\text{b}, \phi_\text{b}).
\end{align*}
$p\bigl(n_{\text{f}_j} \bigl| (\theta_{\text{f}_j}, \phi_{\text{f}_j}) \bigr)$ denotes the marginal probability of observing $n_{\text{f}_j}$ photons, if one measures with angles $\theta_{\text{f}_j}, \phi_{\text{f}_j}$, which is defined as
\begin{align*}
    p\bigl(n_{\text{f}_j} \bigl| (\theta_{\text{f}_j}, \phi_{\text{f}_j}) \bigr)
    =& \int \limits_{0}^{\pi} d\theta_\text{b} \int \limits_{-\pi}^{\pi} d\phi_\text{b} \, p^{(j-1)}(\theta_\text{b},\phi_\text{b}) \times \nonumber \\
    & \times p_\text{t}(N_j, n_{\text{f}_j}, P_{0\text{f}}, P_{1\text{f}}, \theta_{\text{f}_j}, \phi_{\text{f}_j}, \theta_\text{b}, \phi_\text{b}).
\end{align*}
The information gain through the $j$-th experiment is given by the utility function $U(n_{\text{f}_j}, (\theta_{\text{f}_j}, \phi_{\text{f}_j}) )$, which is the difference of the Shannon entropies between the posterior and the prior probability distributions:
\begin{align*}
    &U\bigl(n_{\text{f}_j}, (\theta_{\text{f}_j}, \phi_{\text{f}_j}) \bigr) \nonumber \\
    =& \int \limits_{0}^{\pi} d\theta_\text{b} \int \limits_{-\pi}^{\pi} d\phi_\text{b} \,p\bigl((\theta_\text{b}, \phi_\text{b}) \bigl| n_{\text{f}_j}, (\theta_{\text{f}_j}, \phi_{\text{f}_j}) \bigr)  \times \nonumber \\
    & \phantom{\int \limits_{0}^{\pi} d\theta_\text{b} \int \limits_{-\pi}^{\pi} d\phi_\text{b} \,} \times  \ln\Bigl( p\bigl((\theta_\text{b}, \phi_\text{b}) \bigl| n_{\text{f}_j}, (\theta_{\text{f}_j}, \phi_{\text{f}_j}) \bigr)  \Bigr)  \nonumber \\
    & -\int \limits_{0}^{\pi} d\theta_\text{b} \int \limits_{-\pi}^{\pi} d\phi_\text{b} \,p^{(j-1)}(\theta_\text{b}, \phi_\text{b}) \, \ln\Bigl( p^{(j-1)}(\theta_\text{b}, \phi_\text{b}) \Bigr).
\end{align*}
Averaging over all possible outcomes, $n_{\text{f}_j}$ provides a quantity independent of the hitherto unknown measurement result:
\begin{align*}
    \overline{U}(\theta_{\text{f}_j}, \phi_{\text{f}_j}) =& \sum_{n_{\text{f}_j} = 0}^{N_j} U\bigl(n_{\text{f}_j}, (\theta_{\text{f}_j}, \phi_{\text{f}_j}) \bigr) \, p\bigl(n_{\text{f}_j} \bigl| (\theta_{\text{f}_j}, \phi_{\text{f}_j}) \bigr).
\end{align*}
In order to optimize the $j$-th experiment using the knowledge of the prior probability distribution $p^{(j-1)}(\theta_\text{b},\phi_\text{b})$ of the bank angles, the forger determines the optimal measurement angles $\theta_{\text{f}_j}$, $\phi_{\text{f}_j}$ by maximizing $\overline{U}(\theta_{\text{f}_j}, \phi_{\text{f}_j})$. These angles can then be used for the $j$-th measurement.
In this way, the forger obtains a series of probability distributions $p^{(j)}(\theta_\text{b},\phi_\text{b})$ for the bank angles.
The forger gains in each individual measurement the maximum information due to the optimization using the information from the previous results.
After performing $N_\text{m}$ measurements, the forger takes the angles with the maximum value of the probability distribution $p^{(N_\text{m})}(\theta_\text{b},\phi_\text{b})$ as the optimal guess for the angles of the forged token:
\begin{align*}
    \bigl( \theta_\text{f}^{(\text{Ba})}, \phi_\text{f}^{(\text{Ba})} \bigr) = \text{argmax}\left(  p^{(N_\text{m})}(\theta_\text{b},\phi_\text{b}) \right).
\end{align*}

We only consider the case of two and three measurements.
In both cases, there is no need for optimization of the first measurement due to the freedom of choice of the coordinate system.
Before the second measurement, the forger optimizes the angle $\theta_{\text{f}_2}$ using the first outcome.
For three measurements, the angles $\theta_{\text{f}_3}$ and $\phi_{\text{f}_3}$ are subsequently optimized based on the outcomes of the first two measurements.
We denote the average probability that the bank accepts the forged tokens generated by the Bayesian method using two, or three measurements as $\overline{p}_{\text{f}_2}^{(\text{Ba})}$ and $\overline{p}_{\text{f}_3}^{(\text{Ba})}$, respectively.
In our numerical calculations, we perform the search for the optimal measurement angle $\theta_{\text{f}_2}$ after the first measurement using a brute force search with 2000 values in the interval $[0, \pi)$.
In the three measurements scenario, we also determine the optimal measurement angle $\theta_{\text{f}_2}$, $\theta_{\text{f}_3}$, $\phi_{\text{f}_3}$ using a similar brute force scan.
We present the numerical results in Sec.~\ref{sec:token_security}.

Due to similar performance of this method when compared to the Maximum Likelihood Method, we did not perform any experimental verification due to the increased complexity caused by its iterative nature.


\subsection{Optimal Forged Quantum Tokens}

Now we present the forger's strategy to generate quantum tokens that show the highest possible acceptance rate by the bank after one, two or three measurements.
For this, the forger needs to know the parameters $P_{0\text{b}}$, $P_{1\text{b}}$ of the bank setup and the acceptance threshold $n_\text{T}$ of the bank.
If this knowledge is not available to the forger, the following methods are not applicable.
This shows the advantage of keeping the quantum token implementation of the bank setup secret.

\subsubsection{One Measurement}

Firstly, we consider the case that the forger performs one measurement on all of the $N_1=N$ qubits of the quantum token using the measurement angles $\theta_{\text{f}_1}$, $\phi_{\text{f}_1}$ and detecting  $n_{\text{f}_1}$ photons.
Using this result, the forger has to determine the angles $\theta_\text{f}$, $\phi_\text{f}$ of the forged token in such a way that the average probability
\begin{align}	
	\overline{p}_{\text{f}_1} \bigl((\theta_\text{f}, \phi_\text{f}) \bigl| n_{\text{f}_1} \bigr) =& \int \limits_{0}^{\pi} d\theta_\text{b} \int \limits_{-\pi}^{\pi}  d\phi_\text{b} \, f_\theta(\theta_\text{b}) \, f_\phi(\phi_\text{b}) \times \nonumber \\
	& \times
	p_\text{t}(N_1, n_{\text{f}_1} ,P_{0\text{f}}, P_{1\text{f}}, 0, 0, \theta_\text{b} , \phi_\text{b} ) \times \nonumber \\
	& \times \sum \limits_{n=0}^{n_\text{T}} \, p_\text{t}(N, n, P_{0\text{b}}, P_{1\text{b}},\theta_\text{f}, \phi_\text{f}, \theta_\text{b} , \phi_\text{b} )
	\label{equ:pbaf_1_opt}
\end{align}
of the bank accepting this forged token is maximum.
Since the measurement is performed at $\theta_{\text{f}_1}=0$, the above probability does not depend on $\phi_\text{f}$, so that the forger may set $\phi_\text{f}=0$ and we can write $\overline{p}_{\text{f}_1} (\theta_\text{f}  | n_{\text{f}_1} )$.
The optimal value of the angle
\begin{align}
    \theta_\text{f}^{(\text{opt})} = \text{argmax} \bigl( \overline{p}_{\text{f}_1} (\theta_\text{f}  | n_{\text{f}_1} ) \bigr)
    \label{equ:theta_f_opt_1measurement}
\end{align}
for the forged token must be obtained by brute force numerical methods.
The average probability for the bank accepting this forged token can be calculated by
\begin{align*}
    \overline{p}_{\text{f}_1}^{(\text{opt})} = \sum_{n_{\text{f}_1}=0}^{N_1} \overline{p}_{\text{f}_1} \bigl( \theta_\text{f}^{(\text{opt})}  \bigl| n_{\text{f}_1} \bigr).
\end{align*}

\begin{figure}[tbp!]
    \centering
    \includegraphics[width=\columnwidth]{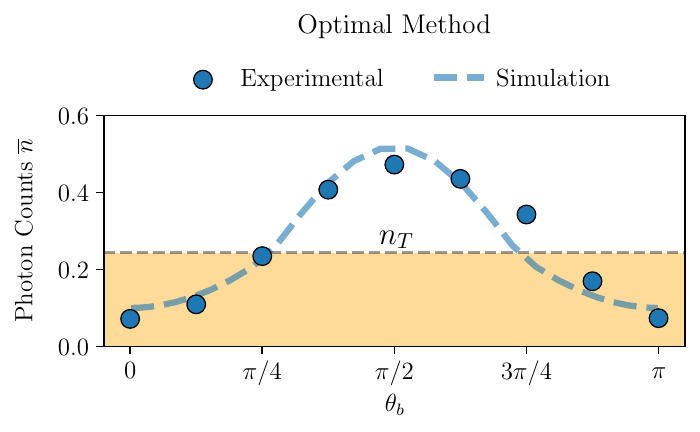}
    \caption{Experimental and simulated normalized counts measured by the bank from a token forged with the optimal method for one measurement axis. As for the other methods, the counts are higher at $\theta_\text{b}=\pi/2$, with  strong agreement between simulation and experimental data from IBMQ Brisbane. The resulting acceptance probability is $\overline{p}_{\text{f}_1}^{(\text{opte})}=0.4937$.}
    \label{fig:optimal_method}
\end{figure}

The optimal method for 1 measurement was experimentally measured with IBMQ Brisbane, as shown in Fig.~\ref{fig:optimal_method}. As in DIT and ML, the method has better chances of success at the poles than at the equator, also showing a great correspondence to the numerical results. Overall, the resulting acceptance probability is $\overline{p}_{\text{f}_1}^{(\text{opte})}=0.4937$.

In order to numerically derive the average acceptance probability of the forged tokens $\overline{p}_{\text{f}_1}^{(\text{opt})}$, we calculate the optimal angle $\theta_\text{f}^{(\text{opt})}$ from Eq.~\eqref{equ:theta_f_opt_1measurement} using Newton's method \cite{Schwarz2011}, starting with an initial guess obtained by a brute force scan of all $\theta$ angles.
In detail, we deployed the Newton's method for searching for $\theta_\text{f}$ such that the derivative of $\overline{p}_{\text{f}_1}^{(\text{opt})}$ with respect to $\theta_\text{f}$ is zero. 
Towards this goal, we analytically derive the second derivatives with respect to $\theta_\text{f}$ and obtained for $N=300$ with Brisbane $\overline{p}_{\text{f}_1}^{(\text{opt})} = 0.3258$, which is lower than the experimental value due to the lower acceptance threshold $n_\text{T}$ in the numerical calculations.
All numerical results are presented and discussed in Sec.~\ref{sec:token_security}.


\subsubsection{Two Measurements}

Precisely determining both token angles with a single measurement is in general not possible. 
Thus, splitting the quantum token into two parts is a more robust attack scenario.
The first part contains $N_1\geq 1$ qubits and the second part $N_2=N-N_1 \geq 1$ qubits.
Due to the freedom choice of the coordinate system, the forger again performs the first measurement with $\theta_{\text{f}_1}=0$ and $\phi_{\text{f}_1}=0$, detecting $n_{\text{f}_1}$ photons.
In the second measurement, the forger uses the angles $\theta_{\text{f}_2}\notin \{0,\pi\}$ and $\phi_{\text{f}_2}=0$, detecting $n_{\text{f}_2}$ photons.
Remember, if $\theta_{\text{f}_2}\in \{0,\pi\}$, then the second measurement is equivalent to the first one.
We want to point out that the forger may choose $\theta_{\text{f}_2}$ depending on the first measurement result $n_{\text{f}_1}$.
Using both measurement results $n_{\text{f}_1}$ and $n_{\text{f}_2}$, the forger prepares the angles to $\theta_\text{f},\phi_\text{f}$ of the forged token.
The average probability that the bank accepts these forged tokens, generated from the two measurement results $n_{\text{f}_1}$ and $n_{\text{f}_2}$, is given by
\begin{align*}
	& \overline{p}_{\text{f}_2} \bigl((\theta_\text{f},\phi_\text{f}) \bigl| N_1, n_{\text{f}_1}, \theta_{\text{f}_2} , n_{\text{f}_2} \bigr) \nonumber \\
	=& \int \limits_{0}^{\pi} d\theta_\text{b} \int \limits_{-\pi}^{\pi}  d\phi_\text{b} \, f_\theta(\theta_\text{b}) \, f_\phi(\phi_\text{b}) \times \nonumber \\
	& \times p_\text{t}(N_1, n_{\text{f}_1}, P_{0\text{f}}, P_{1\text{f}}, 0, 0, \theta_\text{b} , \phi_\text{b} ) \times \nonumber \\
	& \times p_\text{t}(N_2, n_{\text{f}_2}, P_{0\text{f}}, P_{1\text{f}}, \theta_{\text{f}_2}, 0, \theta_\text{b}, \phi_\text{b} ) \times \nonumber \\
	& \times \sum \limits_{n=0}^{n_\text{T}}\,  p_\text{t}(N, n,  P_{0\text{b}}, P_{1\text{b}}, \theta_\text{f}, \phi_\text{f}, \theta_\text{b} , \phi_\text{b} ).
\end{align*}
This scenario has the parameters $N_1$ and $\theta_{\text{f}_2}(n_{\text{f}_1})$ that have to be optimized.
The optimal angles for the forged token are
\begin{align}
    \bigl( \theta_\text{f}^{(\text{opt})}, \phi_\text{f}^{(\text{opt})} \bigr) = \text{argmax} \Bigl( \overline{p}_{\text{f}_2} \bigl((\theta_\text{f},\phi_\text{f}) \bigl| N_1, n_{\text{f}_1}, \theta_{\text{f}_2} , n_{\text{f}_2} \bigr) \Bigr),
    \label{equ:theta_f_opt_phi_f_opt_2measurements}
\end{align}
and must be obtained by brute force numerical methods.
The average probability for the bank to accept these forged tokens can be calculated by
\begin{align*}
    & \overline{p}_{\text{f}_2}^{(\text{opt})} \nonumber \\
    =&\sum_{n_{\text{f}_1}=0}^{N_1} \sum_{n_{\text{f}_2}=0}^{N_2} \overline{p}_{\text{f}_2} \Bigl( \bigl( \theta_\text{f}^{(\text{opt})}, \phi_\text{f}^{(\text{opt})} \bigr) \Bigl| N_1^{(\text{opt})}, n_{\text{f}_1}, \theta_{\text{f}_2}^{(\text{opt})} , n_{\text{f}_2} \Bigr).
\end{align*}
Instead of optimizing the parameters $N_1$ and $\theta_{\text{f}_2}$, we also consider a simpler scenario where $N_1 = N_2 = \frac{N}{2}$ and $\theta_{\text{f}_2}=\frac{\pi}{2}$, so that the measurements are performed in the $z$- and $x$-direction.
We denote the average acceptance probability of the forged tokens for this scenario by $\overline{p}_{\text{f}_2}^{(\text{fix})}$.
In the numerical calculations, we derive the optimal angles $\theta_\text{f}^{(\text{opt})}$,  $\phi_\text{f}^{(\text{opt})}$ from Eq.~\eqref{equ:theta_f_opt_phi_f_opt_2measurements} using the two dimensional Newton's method starting from an initial guess obtained by a brute force scan. 
For this, we derive all necessary derivatives analytically.
In addition, we optimize $N_1$ and the optimal measurement angle $\theta_{\text{f}_2}(n_{\text{f}_1})$ as a function of $n_{\text{f}_1}$ by performing a brute force scan using $N_1=1,2,\ldots,29$ and 2000 points of $\theta_{\text{f}_2}$ in the interval $[0, \pi)$.
We present the optimal acceptance probability of this scenario $\overline{p}_{\text{f}_2}^{(\text{opt})}$ together with the corresponding optimal $N_1$ in Tab.~\ref{tab:hack_token} in Sec.~\ref{sec:token_security}.

\subsubsection{Three Measurements}

Now the forger divides the quantum token into three parts.
The first part contains $N_1\geq 1$ qubits, the second part $N_2\geq 1$ and the third part $N_3=N-N_1-N_2$ qubits.
Again, due to the freedom choice of the coordinate system, the forger performs the first measurement with $\theta_{\text{f}_1}=0$ and $\phi_{\text{f}_1}=0$, detecting $n_{\text{f}_1}$ photons.
In the second measurement, the forger uses the angles $\theta_{\text{f}_2}\notin \{0,\pi\}$ and $\phi_{\text{f}_2}=0$ and detects $n_{\text{f}_2}$ photons.
In the third measurement, the forger uses the angles $\theta_{\text{f}_3}$ and $\phi_{\text{f}_3}$, detecting $n_{\text{f}_3}$ photons.
Using the three measurement results $n_{\text{f}_1}$, $n_{\text{f}_2}$ and $n_{\text{f}_3}$, the forger sets the angles $\theta_\text{f},\phi_\text{f}$ of the forged token.
The average probability that the bank accepts these forged tokens generated from the measurement results $n_{\text{f}_1}$, $n_{\text{f}_2}$ and $n_{\text{f}_3}$ is given by
\begin{align*}
	& \overline{p}_{\text{f}_3} \bigl( (\theta_\text{f},\phi_\text{f}) \bigl | N_1, n_{\text{f}_1}, N_2, \theta_{\text{f}_2}, n_{\text{f}_2}, \theta_{\text{f}_3}, \phi_{\text{f}_3}, n_{\text{f}_3} \bigr) \nonumber \\
	=& \int \limits_{0}^{\pi} d\theta_\text{b} \int \limits_{-\pi}^{\pi}  d\phi_\text{b} \, f_\theta(\theta_\text{b}) \, f_\phi(\phi_\text{b}) \times \nonumber \\
	& \times p_\text{t}(N_1, n_{\text{f}_1}, P_{0\text{f}}, P_{1\text{f}} , 0, 0, \theta_\text{b} , \phi_\text{b} ) \times \nonumber \\
	& \times p_\text{t}(N_2, n_{\text{f}_2}, P_{0\text{f}}, P_{1\text{f}}, \theta_{\text{f}_2}, 0, \theta_\text{b} , \phi_\text{b} ) \times \nonumber \\
	& \times p_\text{t}(N_3, n_{\text{f}_3}, P_{0\text{f}}, P_{1\text{f}}, \theta_{\text{f}_3}, \phi_{\text{f}_3}, \theta_\text{b}, \phi_\text{b} ) \times \nonumber \\
	& \times \sum \limits_{n=0}^{n_\text{T}} p_\text{t}(N, n, P_{0\text{b}}, P_{1\text{b}}, \theta_\text{f}, \phi_\text{f}, \theta_\text{b} , \phi_\text{b} ).
\end{align*}
The measurement scheme in this scenario has the parameters $N_1$, $N_2(n_{\text{f}_1})$, $\theta_{\text{f}_2}(n_{\text{f}_1})$, $\theta_{\text{f}_3}(n_{\text{f}_1},n_{\text{f}_2})$ and $\phi_{\text{f}_3}(n_{\text{f}_1},n_{\text{f}_2})$ that have to be optimized.
Since numerically determining the optimal parameters is quite demanding, we only consider the simpler scenario where the token is divided in equal parts with $N_1=N_2=N_3=\frac{N}{3}$ qubits and the angles $\theta_{\text{f}_2}=\frac{\pi}{2}$, $\phi_{\text{f}_2}= 0$ , $\theta_{\text{f}_3}=\frac{\pi}{2}$, $\phi_{\text{f}_3}= \frac{\pi}{2}$, so that the measurements are performed in the $z$-, $x$- and $y$-direction.
We denote the average probability for the bank to accept the forged tokens in this measurement procedure by $\overline{p}_{\text{f}_3}^{(\text{fix})}$ and present the results in Sec.~\ref{sec:token_security}.


\subsection{Token Security}
\label{sec:token_security}

\begin{figure}[htbp!]
	\includegraphics[width=0.49\textwidth]{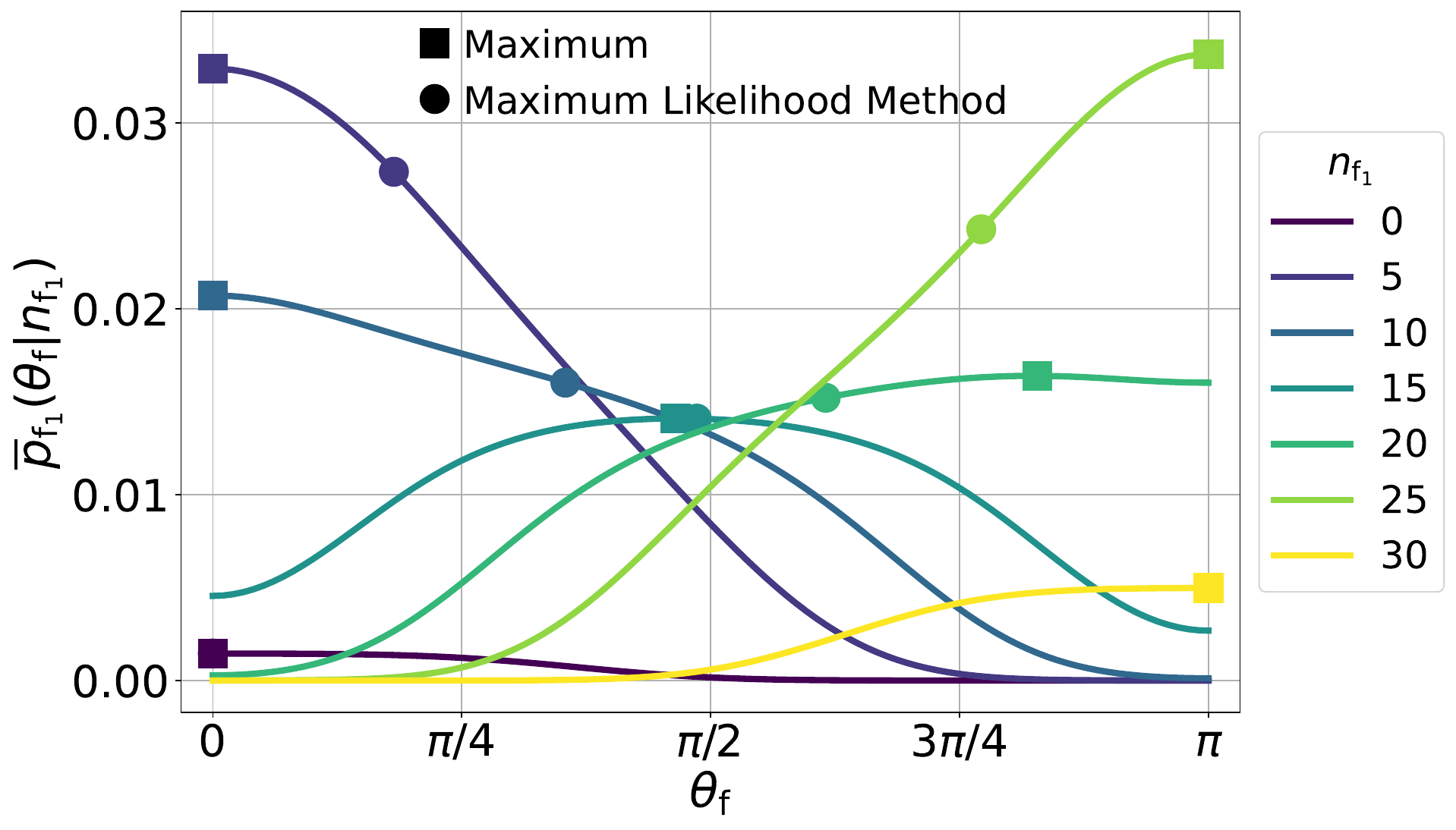}
	\caption{Average acceptance probability $\overline{p}_{\text{f}_1} (\theta_\text{f}  | n_{\text{f}_1} )$ of the bank accepting the forged tokens as a function of $\theta_\text{f}$ for different measurement results $n_{\text{f}_1}$ of one measurement on the quantum token for IBMQ Brisbane with $N=30$ qubits. 
	The squares indicate the maximum and the dots show the results from the maximum likelihood method.
	It is clearly visible that the maximum likelihood method is suboptimal. This can be also deduced from Tab.~\ref{tab:hack_token} where we have compared the different methods with corresponding acceptance probabilities.}
	\label{fig:pbaf_thetaf}
\end{figure}

Now we discuss the numerical results of the different attack methods with regard to the acceptance probability.
We first simulate attacks using one measurement on the whole quantum token.
In Fig.~\ref{fig:pbaf_thetaf} we show the average probability for the bank to accept the forged tokens $\overline{p}_{\text{f}_1} (\theta_\text{f}  | n_{\text{f}_1} )$ from Eq.~\eqref{equ:pbaf_1_opt} for various measurement results $n_{\text{f}_1}$ as a function of $\theta_\text{f}$ derived for the IBMQ Brisbane with $N=30$ qubits.
In addition, we indicate the results of the maximum likelihood method via dots and the optimal solutions as crosses.
One can clearly see that the maximum likelihood method does not provide optimal results in all cases. 
If $n_{\text{f}_1}$ changes, the qubit is with high probability in a different state, thus the probability distribution is modified as can be seen in Fig.~\ref{fig:pbaf_thetaf}. 
The probability distribution becomes broader for $n_{\text{f}_1} \to N/2 = 15$, since the qubit state is more probably initialized in the equatorial plane, where the density of states is higher.

In Tab.~\ref{tab:hack_token}, we present all numerical and experimental results for the average acceptance probability of the differently generated forged tokens for $N=30$ and $N=300$ qubits in the different IBMQs.
One can clearly see, that the acceptance probability for all cloning methods decreases, if one uses $N=300$ instead of $N=30$ qubits.
Here, we observe a higher security level for an increase of qubits in the token for all methods. However, for a very high number of qubits the ensemble will behave classically and the quantum projection noise will vanish in the shot noise, so that the security will decrease again. 
In addition, the results in Tab.~\ref{tab:hack_token} show that the optimization of the parameters in the two measurement case $\overline{p}_{\text{f}_2}^{(\text{opt})}$ do not provide a significant improvement of the acceptance probability compared to the simple case $\overline{p}_{\text{f}_2}^{(\text{fix})}$.

\begin{table*}[tb!]
	\begin{tabular}{|l||r|r|c||c||c|c||c|c|c|c|c||c|c|c|c|}
		\hline
		IBMQ   & $N$ & $n_\text{T}$ & $\overline{p}_\text{b}$ & $\overline{p}_{\text{f}_0}$& $\overline{p}_{\text{f}_1}^{(\text{ML})}$& $\overline{p}_{\text{f}_1}^{(\text{opt})}$&
		$\overline{p}_{\text{f}_2}^{(\text{Ba})}$&
		$\overline{p}_{\text{f}_2}^{(\text{ML})}$& $\overline{p}_{\text{f}_2}^{(\text{fix})}$& $N_1$ & $\overline{p}_{\text{f}_2}^{(\text{opt})}$& $\overline{p}_{\text{f}_3}^{(\text{DIT})}$&
		$\overline{p}_{\text{f}_3}^{(\text{Ba})}$&
		$\overline{p}_{\text{f}_3}^{(\text{ML})}$& $\overline{p}_{\text{f}_3}^{(\text{fix})}$\\ \hline
		Sherb. &  30 &  0 & $\sim$1 & 0.03226 & 0.1393 & 0.1409 & 0.2992 & 0.2969 & 0.3196 & 15 & 0.3196 & 0.1094 & 0.4655 & 0.3474 & 0.4832 \\ 
		Kyiv   &  30 &  0 & $\sim$1 & 0.03370 & 0.1383 & 0.1408 & 0.2923 & 0.2879 & 0.3094 & 15 & 0.3099 & 0.1091 & 0.4496 & 0.3423 & 0.4651 \\ 
		Osaka  &  30 &  5 & 0.99981 & 0.1836  & 0.4169 & 0.4726 & 0.6664 & 0.6797 & 0.7372 & 15 & 0.7372 & 0.3847 & 0.9291 & 0.8592 & 0.9384 \\ 
		Brisb. &  30 & 10 & 0.99991 & 0.3045  & 0.5450 & 0.6475 & 0.7552 & 0.7693 & 0.8537 & 15 & 0.8537 & 0.5183 & 0.9731 & 0.9356 & 0.9776 \\ 
		Kyoto  &  30 & 14 & 0.99986 & 0.5128  & 0.7077 & 0.8367 & 0.8260 & 0.8454 & 0.9112 & 14 & 0.9192 & 0.6707 & 0.9546 & 0.9299 & 0.9583 \\ 
		\hline
		Sherb. & 300 &  0 & $\sim$1 & 0.003322& 0.04268 & 0.04403 & 0.2780 & 0.2299 & 0.2815 & * & * & 0.04283 & * & 0.3357 & 0.5589 \\ 
		Kyiv   & 300 &  0 & $\sim$1 & 0.003471& 0.04157 & 0.04273 & 0.2594 & 0.2165 & 0.2630 & * & * & 0.03775 & * & 0.3139 & 0.5299 \\ 
		Osaka  & 300 & 20 & 0.99982 & 0.04586 & 0.2094  & 0.2307  & 0.5863 & 0.5945 & 0.6326 & * & * & 0.1962  & * & 0.9473 & 0.9954 \\
		Brisb. & 300 & 50 & 0.99986 & 0.08271 & 0.2835  & 0.3258  & 0.6340 & 0.6404 & 0.7104 & * & * & 0.2770  & * & 0.9803 & 0.9990 \\
		Kyoto  & 300 & 83 & 0.99987 & 0.1535  & 0.3869  & 0.4591  & 0.6792 & 0.6905 & 0.7812 & * & * & 0.3757  & * & 0.9758 & 0.9982 \\
		\hline
	\end{tabular}
	\begin{tabular}{|l||c|c|c|c|c|c|c|}
		\hline
		IBMQ & $N$ & $n_\text{T}^\text{(e)}$ & $\overline{p}_{\text{f}_1}^{(\text{MLe})}$ & $\overline{p}_{\text{f}_1}^{(\text{opte})}$ &
		$\overline{p}_{\text{f}_2}^{(\text{MLe})}$&   $\overline{p}_{\text{f}_3}^{(\text{DITe})}$&
		$\overline{p}_{\text{f}_3}^{(\text{MLe})}$ \\ \hline
		Brisbane & 300 & 73 & 0.4698 & 0.4937 & 0.6688 & 0.3769 & 0.9847 \\ \hline
	\end{tabular}
	\caption{(top) Average acceptance probability of the bank and forged tokens generated from various attack scenarios for $N=30$ and $N=300$ qubits in a quantum token on the five IBMQs with $P_{0\text{b}}=P_{0\text{f}}$ and $P_{1\text{b}}=P_{1\text{f}}$. For $N=300$, the stars indicate the attack scenarios that were not simulated due to the very high computational cost. Note that the numerical brute force methods with $\overline{p}_{\text{f}_1}^{(\text{opt})}$, $\overline{p}_{\text{f}_2}^{(\text{opt})}$ and $\overline{p}_{\text{f}_3}^{(\text{fix})}$ achieve always larger acceptance values than the state estimation methods with biggest advantages for three measurements. The values are also visualized in Fig.~\ref{fig:acceptance_probabilites}.\\
	(bottom) For direct comparison we have repeated the experimental values for the Brisbane IBMQ with $N=300$.}
	\label{tab:hack_token}
\end{table*}

\begin{figure}[htb!]
    \centering
    \includegraphics[width=\columnwidth]{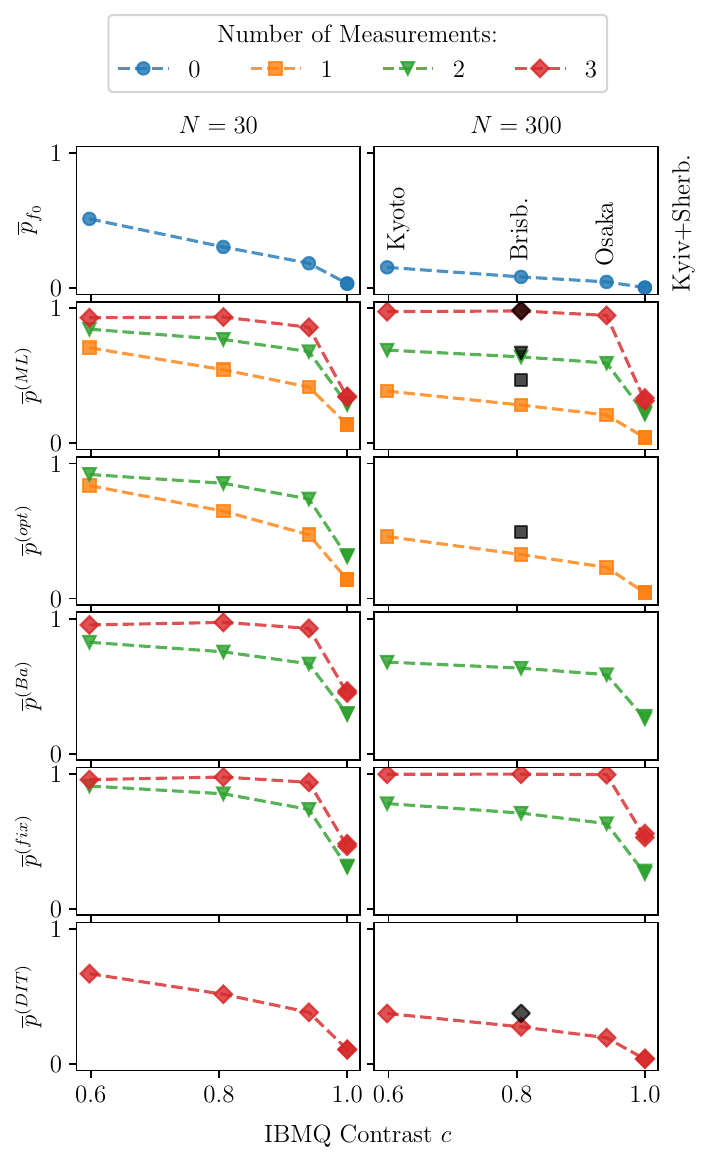}
\caption{ Acceptance probabilities from Tab.~\ref{tab:hack_token} for the quantum token as implemented on the five different IBMQ platforms as a function of the different contrasts $c$ from Eq.~\eqref{equ:contrast_C}. Black symbols denote experimental verification with the corresponding number of measurements (see legend). It is obvious that with better contrast more forged tokens are rejected by the bank. It can be clearly seen that three measurements are always advantageous when preparing a forged token. With better contrast this advantage is reduced.}
    \label{fig:acceptance_probabilites}
\end{figure}

As one can clearly see in Tab.~\ref{tab:hack_token} and also in Fig.~\ref{fig:acceptance_probabilites}, the ability to perform measurements on sub-ensembles of the quantum token provides significantly better acceptance probabilities for the forged tokens.
The direct inversion method provides the worst results even using three measurements.
The Bayesian method provides significantly better results than the maximum likelihood method for three measurements.
For two measurements, both methods show almost similar performance.
For the one-, two- and three measurement case, the optimal scenario introduced in this work always outperforms the other methods presented.
Nevertheless, as can be seen in Tab.~\ref{tab:hack_token}, the acceptance probability $\overline{p}_\text{b}$ of the token generated by the bank is always higher than for the forged tokens $\overline{p}_\text{f}$ for the different methods. We also summarized the experimental acceptance probabilities for the Brisbane IBMQ in the table. In the comparison it might be striking that the experimental acceptance probabilities for the single measurements deviate from our simulations for the same platform. This is caused by a different adjustment of the $n_\text{T}^\text{(e)}$ in order to achieve the same probability $\overline{p}_\text{b}>0.9998$ for the bank accepting its own generated tokens. This was necessary because in the real experiments additional errors such as gate errors and decoherence effects needed to be taken into account which were not present in our simulations. 

Observe that a small improvement of the quantum hardware, i.e. increase of contrast $c$, can induce a great improvement of the security, as can be seen by comparing Osaka with Kyiv and Sherbrooke. 
The security of our protocol thus will benefit from the evolution of the quantum hardware.

\section{Quantum Coin Consisting of Quantum Tokens}
\label{sec:coin}

As exemplified with the experimental results from Brisbane an increase in $n_\text{T}^\text{(e)}$ was necessary to keep $\overline{p}_\text{b}>0.9998$ valid. This is carried through at the cost of making it easier for the forger to generate forged tokens. In order to achieve a predefined security of the protocol we assume that $M$ quantum tokens with prepared with individual angles are combined into a quantum coin.
There are two conditions for the coin, which should be fulfilled:
\begin{itemize}
    \item[1.] The probability for the bank rejecting their own generated coins is less than a given limit $\varepsilon_\text{cb}>0$.
    \item[2.] The probability for the bank to accept forged coins is less than a given limit $\varepsilon_\text{cf}>0$.
\end{itemize}
The usage of several quantum tokens within a coin allows the bank to generate coins that fulfill both conditions for any given limits $\varepsilon_\text{cb}>0$ and $\varepsilon_\text{cf}>0$, if the average acceptance probability of the bank generated tokens $p_\text{b}$ is bigger than the acceptance probability of forged tokens $p_\text{f}$, which is fulfilled for all cases and attack scenarios, as can be seen in Tab.~\ref{tab:hack_token}.

The bank validates a coin, if at least $n_\text{cT}$ tokens within the coin are accepted.
The acceptance threshold $n_\text{cT}$ is determined from the condition 1.:
\begin{align}
    \sum_{n=n_\text{cT}}^{M} p_\text{b}^n \, (1-p_\text{b})^{M-n} < \varepsilon_\text{cb}.
    \label{equ:condition_epsilon_cb}
\end{align}
Here we use the fact that the probability for the bank accepting $n$ quantum tokens within a coin is given by a binomial distribution with probability $p_\text{b}$.
Furthermore, if the average forged token acceptance probability $p_\text{f}$ is given, the bank demands to have an acceptance rate of forged coins being less than $\varepsilon_\text{cf}$:
\begin{align}
    \sum_{n=0}^{n_\text{cT}-1} p_\text{f}^n \, (1-p_\text{f})^{M-n} < \varepsilon_\text{cf}.
    \label{equ:condition_epsilon_cf}
\end{align}
In Appendix~\ref{sec:safety_proof}, we proof that for any $\varepsilon_\text{cb}>0$ and $\varepsilon_\text{cf}>0$ a number $M$ of quantum tokens in the coin can be found such that conditions 1 and 2 are fulfilled. 
Practically, one can perform the following iterative procedure:
One starts with a given $M$ and derives the corresponding $n_\text{cT}$ from Eq.~\eqref{equ:condition_epsilon_cb}.
Then one checks if condition \eqref{equ:condition_epsilon_cf} is fulfilled. If not, one start the procedure again with $M+1$.

Finally, we simulate a quantum coin composed of a variable number $M$ of quantum tokens.
For each number $M$, we calculate the acceptance threshold of the coin $n_\text{cT}$ using Eq.~\eqref{equ:condition_epsilon_cb} in such a way that the acceptance probability of the bank generated coins is always bigger than $0.99999$ or in other words, $\varepsilon_\text{cb}<0.00001$.
Then, we consider for zero, one, two and three measurements the optimal forged token scenario from Tab.~\ref{tab:hack_token} and derive the average acceptance probability of the corresponding forged coin using Eq.~\eqref{equ:condition_epsilon_cb}.
Note, we perform this calculations with 100 digits numerical precision.
We present the obtained results for Brisbane in Tab.~\ref{tab:hack_coin_brisbane} and visualize the values in Fig.~\ref{fig:acceptance_coin_Brisbane}.
In Tab.~\ref{tab:hack_coin_all} of the appendix, we show the results for all IBMQ platforms.
One can clearly see that for the coin any level of security can be obtained by just increasing the number of quantum tokens.

\begin{figure}[htb!]
    \centering
    \includegraphics[width=\columnwidth]{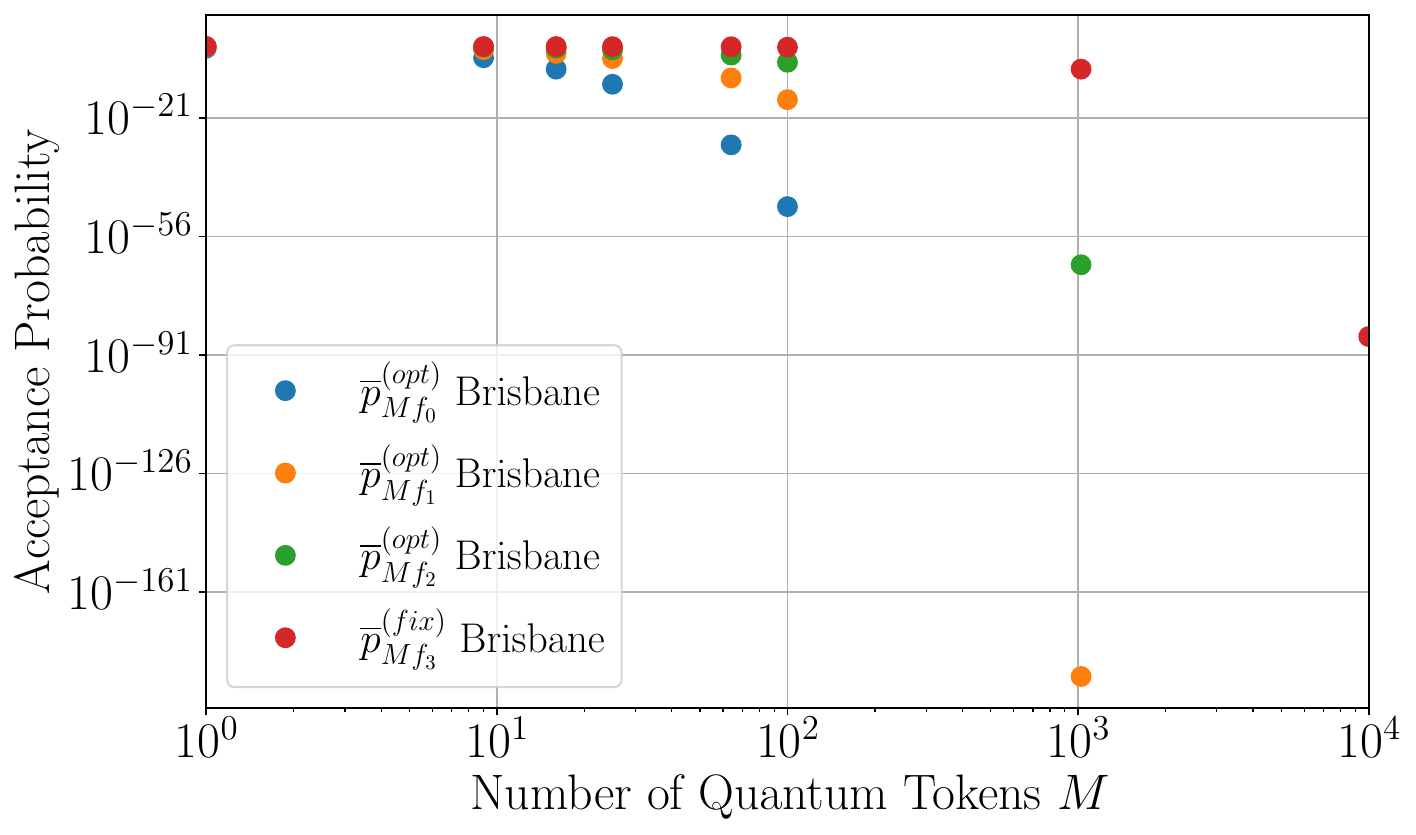}
\caption{Acceptance probabilities from Tab.~\ref{tab:hack_coin_brisbane} for Brisbane are shown for the different hack scenarios as a function of the number of quantum tokens $M$.}
    \label{fig:acceptance_coin_Brisbane}
\end{figure}

\begin{table*}[htb!]
\begin{tabular}{|l|r||r||c|c|c|c|}
	\hline
	IBMQ & $M$ & $n_\text{cT}$ & $\overline{p}_{\text{Mf}_0}$ & $\overline{p}_{\text{Mf}_1}^{(\text{opt})}$ & $\overline{p}_{\text{Mf}_2}^{(\text{opt})}$ & $\overline{p}_{\text{Mf}_3}^{(\text{fix})}$ \\ 
	\hline
	          &     1 &     1 & 0.3045 & 0.6475 & 0.8537 & 0.9776 \\
	          &     9 &     8 & 4.851$\cdot10^{-4}$ & 0.1180 & 0.6123 & 0.9837 \\
	          &    16 &    15 & 2.051$\cdot10^{-7}$ & 9.270$\cdot10^{-3}$ & 0.2978 & 0.9511 \\
	          &    25 &    24 & 7.143$\cdot10^{-12}$ & 2.790$\cdot10^{-4}$ & 0.1013 & 0.8927 \\
	Brisbane  &    64 &    62 & 9.496$\cdot10^{-30}$ & 5.260$\cdot10^{-10}$ & 2.857$\cdot10^{-3}$ & 0.8269 \\
	          &   100 &    98 & 5.951$\cdot10^{-48}$ & 2.025$\cdot10^{-16}$ & 2.208$\cdot10^{-5}$ & 0.6113 \\
	          &  1024 &  1021 & 3.323$\cdot 10^{-520}$ & 1.483$\cdot 10^{-186}$ & 4.143$\cdot10^{-65}$ & 2.058$\cdot10^{-7}$ \\
	          & 10000 &  9993 & 4.794$\cdot 10^{-5138}$ & 7.021$\cdot 10^{-1866}$ & 9.746$\cdot 10^{-669}$ & 2.771$\cdot10^{-86}$ \\
	 \hline
\end{tabular}
\caption{Acceptance probability for forged coins on IBMQ Brisbane containing 1, 9, 100, 1024, 10000 quantum tokens, each one with $N=30$ qubits. Here, the acceptance probability of the bank coins with more than 1 token is always bigger than 0.99999, where we take 100 digits of precision.}
\label{tab:hack_coin_brisbane}
\end{table*}

\section{Discussion and Outlook}
\label{sec:discussion}

We have presented an ensemble-based quantum token protocol with quantum coins consisting of individual quantum tokens, each one containing an ensemble of qubits prepared in the same state. We have shown attack scenarios, which describe how to measure the state of the quantum tokens with quantum state tomography, Bayesian method and the Maximum Likelihood Method using one, two or three measurements. We have shown the optimal scheme to generate forged tokens with highest acceptance probability of the bank. Remarkably, these schemes provides significantly better results than the state-of the art quantum tomography methods. 
This is due to the fact that optimal state estimation has a different objective than trying to trick the bank into accepting a forged token when the attacker has knowledge about the setup parameters of the bank. Finally, we have shown that the coin becomes arbitrary safe if the number of quantum tokens within the coin is increased.

The presented protocol is hardware agnostic and can be applied to any qubit for which ensemble initialization, manipulation and readout is feasible. The ensemble-based quantum token protocol must be fortified against fake tokens. As an example, a forger might generate a dark token that always delivers zero photons. As the bank would normally rotate all states back into the dark state, such a dark token would be always accepted. As a countermeasure the bank should arbitrarily select tokens which are measured in the bright state, such that a forger cannot guess the dark and bright tokens.

Even though the IBMQ presented itself as an excellent platform for a hardware agnostic benchmark of the quantum coin protocol and testing of the attack scenarios, superconducting architectures have severe limitations regarding their applications for a quantum coin device. Foremost, superconduting qubits can hardly be made mobile, due to low temperature constrains. Optimal platforms for implementing the presented ensemble-based quantum token protocol have to further rely on long qubit storage times. Therefore, nuclear spin qubits are preferable as storage qubits. Hybrid quantum systems such as NV-color centers coupled to nuclear spin qubits or cold Alkali Atoms with long lived hyperfine splitted ground states would provide the ideal platform for an implementation of a real-world quantum token. The diamond platform can be used at room temperature and has reached $T_2$ lifetimes of 90 seconds with Floquet prethermalized nuclear spins \cite{Beatrez2021}. It is additionally attractive due to the possibility of miniaturizing an entire diamond quantum coin on a single diamond substrate using nanofabrication techniques that allow to use diamond nano-pillars with integrated NV-centers \cite{Schmidt2019} as a quantum token. This technique has the additional benefit that performing measurements of sub-ensembles of the quantum token may be unfeasible, thus greatly limiting the attacker's ability to forge fake tokens. Optimization of fabrication methods for quantum tokens based on color centers in diamond \cite{token_fabrication}, as well as qubit control techniques for fast state transfer between solid state spins in these systems \cite{fast_swap} could provide a robust room temperature platform for implementing ensemble-based quantum tokens. 

\begin{acknowledgments}
	This work was supported by the German Federal Ministry of Education and Research (BMBF) within the initiative "Grand Challenge of Quantumcommunication" under the project "\textit{DIamant-basiert QuantenTOKen}" (DIQTOK - n\textsuperscript{o} 16KISQ034) and the German Science foundation (DFG, grant 410866378).
	Computations were performed on the IT Servicecenter (ITS) University of Kassel and on the computing cluster FUCHS University of Frankfurt.
	We thank Janis Nötzel from the TUM School of Computation, Information and Technology from Munich, Manika Bhardwaj and Moritz Göb for fruitful discussions.
\end{acknowledgments}

\clearpage
\appendix

\section{Safety proof of ensemble-based quantum token protocol}
\label{sec:safety_proof}

We define the hypothesis $H_0$ and $H_1$ as follows:

\begin{description}
    \item[$H_0$] Bank prepared the coin.
    \item[$H_1$] Forger prepared the coin.
\end{description}

We have the following errors $\varepsilon_\text{cb},\varepsilon_\text{cf} \in [0,1]$ of first and second kind:

\begin{description}
    \item[Error of first kind $\varepsilon_\text{cb}$] Bank declines coin even thought it was prepared by the bank.
    \item[Error of second kind $\varepsilon_\text{cf}$] Bank accepts coin even thought it was prepared by the forger.
\end{description}

We have $M$ quantum tokens in the coin.
If $H_0$ is valid, then each token is represented by the random variable $X_j\in\{0,1\}$, which is 1, if the token is  accepted by the bank, which has the probability $p_\text{b}$, and 0 otherwise.
If $H_1$ is valid, then each token is represented by the random variable $Y_j\in\{0,1\}$, which is 1, if the token is accepted by the bank, which has the probability $p_\text{f}$, and 0 otherwise.
We define
\begin{align*}
    \langle X\rangle :=& \frac{1}{M} \sum_{j=1}^{M} X_j, \\
    \langle Y\rangle :=& \frac{1}{M} \sum_{j=1}^{M} Y_j.
\end{align*}
The set of quantum tokens have to fulfill the following two conditions:
\begin{align*}
    \langle X\rangle \geq& \frac{n_\text{cT}}{M}, \\
    \langle Y\rangle <&  \frac{n_\text{cT}}{M}.
\end{align*}
The number $n_\text{cT}$, which corresponds to the accept criteria, is well chosen, if the following two conditions hold:
\begin{align*}
    p\left(\langle X\rangle < \frac{n_\text{cT}}{M}\right) \leq & \varepsilon_\text{cb}, \\
    p\left(\langle Y\rangle \geq \frac{n_\text{cT}}{M}\right) \leq & \varepsilon_\text{cf} \ \Longleftrightarrow \ p\left(\langle Y\rangle < \frac{n_\text{cT}}{M}\right) \geq 1 - \varepsilon_\text{cf}.
\end{align*}
The random variables $X_j$, $Y_j$ obey a Bernoulli distribution with mean $p_\text{b}, p_\text{f}$ and standard deviation $\sqrt{p_\text{b}\, (1-p_\text{b})}, \sqrt{p_\text{f}\, (1-p_\text{f})}$, respectively.
The central limit theorem \cite{Handl2018} yields that
\begin{align*}
    \lim \limits_{M \to \infty} p\left(\sqrt{M} \, \frac{\langle X \rangle - p_\text{b}}{\sqrt{p_\text{b}\, (1-p_\text{b})}} \leq z \right) =& \Phi(z), \\
    \lim \limits_{M \to \infty} p\left(\sqrt{M} \, \frac{\langle Y \rangle - p_\text{f}}{\sqrt{p_\text{f}\, (1-p_\text{f})}} \leq z \right) =& \Phi(z),
\end{align*}
where the $\Phi(z)$ is the cumulative distribution function of the normal distribution and hence given as
\begin{equation*}
    \Phi(z) = \int \limits_{-\infty}^{z} \frac{1}{\sqrt{2\pi}} e^{-\frac{x^2}{2}} \, dx = \frac{1}{2}\left( 1 + \text{erf} \left(\frac{z}{\sqrt{2}}\right)\right).
\end{equation*}
If we express the above limit for finite $M$, we obtain for each $\varepsilon >0$ exists a $n_\text{b}\in \mathbb{N}$ so that for all $M \geq n_\text{b}$:
\begin{align*}
    p\left(\sqrt{M} \, \frac{\langle X \rangle - p_\text{b}}{\sqrt{p_\text{b}\, (1-p_\text{b})}} \leq z \right) \in &  \Bigl(\Phi(z) - \varepsilon, \Phi(z) + \varepsilon \Bigr), \\
    p\left(\sqrt{M} \, \frac{\langle Y \rangle - p_\text{f}}{\sqrt{p_\text{f}\, (1-p_\text{f})}} \leq z \right) \in &  \Bigl(\Phi(z) - \varepsilon, \Phi(z) + \varepsilon \Bigr).
\end{align*}
Using the error bounds $\varepsilon_\text{cb},\varepsilon_\text{cf}$ for the errors of first and second kind, we obtain finally
\begin{align*}
    \varepsilon_\text{cb} \geq& \Phi(z_X) + \varepsilon \nonumber \\
     \geq& p\left(\sqrt{M} \, \frac{\langle X\rangle - p_\text{b}}{\sqrt{p_\text{b}\, (1-p_\text{b})}} < \underbrace{\sqrt{M} \, \frac{\frac{n_\text{cT}}{M} - p_\text{b}}{\sqrt{p_\text{b}\, (1-p_\text{b})}}}_{=:z_X} \right) \nonumber \\
    = & p\left(\langle X\rangle < \frac{n_\text{cT}}{M}\right)
\end{align*}
and
\begin{align*}
    1-\varepsilon_\text{cf} \leq& \Phi(z_Y) - \varepsilon \nonumber \\
    \leq& p\left(\sqrt{M} \, \frac{\langle Y\rangle - p_\text{f}}{\sqrt{p_\text{f}\, (1-p_\text{f})}} < \underbrace{\sqrt{M} \, \frac{\frac{n_\text{cT}}{M} - p_\text{f}}{\sqrt{p_\text{f}\, (1-p_\text{f})}}}_{=:z_Y} \right)  \nonumber \\
    =& p\left(\langle Y\rangle < \frac{n_\text{cT}}{M}\right).
\end{align*}
It follows
\begin{align*}
    \Phi(z_X) \leq& \varepsilon_\text{cb} - \varepsilon, \\
    \Phi(z_Y) \geq& 1-\varepsilon_\text{cf} + \varepsilon,
\end{align*}
which is equivalent to
\begin{align*}
    \sqrt{M} \, \frac{\frac{n_\text{cT}}{M} - p_\text{b}}{\sqrt{p_\text{b}\, (1-p_\text{b})}} \leq & \Phi^{-1}(\varepsilon_\text{cb}-\varepsilon), \\
    \sqrt{M} \, \frac{\frac{n_\text{cT}}{M} - p_\text{f}}{\sqrt{p_\text{f}\, (1-p_\text{f})}} \geq & \Phi^{-1}(1-\varepsilon_\text{cf} + \varepsilon),
\end{align*}
since $\Phi(z)$ is a monotonously increasing function. We obtain further
\begin{align}
    n_\text{cT} \leq & \Phi^{-1}(\varepsilon_\text{cb}-\varepsilon)  \sqrt{p_\text{b}\, (1-p_\text{b})} \, \sqrt{M} + p_\text{b}\, M,
    \label{equ:n_bt_upper_bound}\\
    n_\text{cT} \geq & \underbrace{\Phi^{-1}(1-\varepsilon_\text{cf} + \varepsilon)}_{=-\Phi^{-1}(\varepsilon_\text{cf}-\varepsilon)} \, \sqrt{p_\text{f}\, (1-p_\text{f})} \, \sqrt{M} + p_\text{f}\, M.
    \label{equ:n_bt_lower_bound}
\end{align}
In summary, the above equations yield the interval for $n_\text{cT}$ given by the boundaries above.
Thus, $n_\text{cT}$ is well chosen, if there exists a natural number within this interval.
This is always the case, if the length of this interval is larger or equal to 1.
Therefore, we obtain the following sufficient condition for $n_\text{cT}$ to be well chosen:
\begin{align*}
    1 \leq& \Phi^{-1}(\varepsilon_\text{cb}-\varepsilon)  \sqrt{p_\text{b}\, (1-p_\text{b})} \, \sqrt{M} + p_\text{b}\, M + \nonumber \\
    & +\Phi^{-1}(\varepsilon_\text{cf}-\varepsilon) \, \sqrt{p_\text{f}\, (1-p_\text{f})} \, \sqrt{M} - p_\text{f}\, M
\end{align*}
This tranforms to
\begin{align*}
     & \underbrace{-\Phi^{-1}(\varepsilon_\text{cb}-\varepsilon)  \sqrt{p_\text{b}\, (1-p_\text{b})} - \Phi^{-1}(\varepsilon_\text{cf}-\varepsilon) \, \sqrt{p_\text{f}\, (1-p_\text{f})}}_{=:d} \nonumber \\
    \leq & \underbrace{(p_\text{b} - p_\text{f})}_{=:\Delta p} \, \sqrt{M} - \frac{1}{\sqrt{M}}.
\end{align*}
The left hand side is bigger than zero due to $\varepsilon_\text{cb}-\varepsilon,\varepsilon_\text{cf}-\varepsilon \leq \frac{1}{2}$.
If we have $p_\text{b} > p_\text{f}$, then the right hand side tends to $\infty$ with $M\to \infty$.
This means, if $M$ is large enough, there will always exist a well chosen $n_\text{b}$.
We obtain further
\begin{align*}
                             0\leq&   \Delta p \, \sqrt{M} - \frac{1}{\sqrt{M}} - d \nonumber \\
\Leftrightarrow \qquad   0\leq&  M - \frac{d}{\Delta p} \, \sqrt{M} - \frac{1}{\Delta p} + \underbrace{\frac{d^2}{4\,\Delta p^2} - \frac{d^2}{4\,\Delta p^2}}_{=0} \nonumber
\end{align*}
and finally
\begin{align*}
\Leftrightarrow \qquad  \frac{4\, \Delta p + d^2}{4\, \Delta p^2} \leq& \left(\sqrt{M} - \frac{d}{2\, \Delta p}\right)^2 \nonumber \\
\Leftrightarrow \qquad  \frac{\sqrt{4\, \Delta p + d^2}}{2\, \Delta p} \leq& \left|\sqrt{M} - \frac{d}{2\, \Delta p}\right|.
\end{align*}
If we have $\sqrt{M} - \frac{d}{2\, \Delta p} < 0$, then the following is fulfilled
\begin{align*}
    \sqrt{M} \leq& \frac{d-\sqrt{4\, \Delta p + d^2}}{2\, \Delta p} < 0.
\end{align*}
This means that we have in this case no solution for $M$, since $\sqrt{M}\geq 0$.
If we have in the other case $\sqrt{M} - \frac{d}{2\, \Delta p} \geq 0$, then it follows
\begin{align*}
    \sqrt{M} \geq \frac{d+\sqrt{4\, \Delta p + d^2}}{2\, \Delta p}.
\end{align*}
If the condition above is valid, then automatically  $\sqrt{M} - \frac{d}{2\, \Delta p} \geq 0$ is fulfilled.
Consequently, the only solution for $M$ is given by
\begin{align*}
    M \geq \max \left\{ \left(\frac{d+\sqrt{4\, \Delta p + d^2}}{2\, \Delta p}\right)^2, n_\text{b} \right\}.
\end{align*}
After determining of $M$, one obtains $n_\text{cT}$ from Eq. \eqref{equ:n_bt_upper_bound} and Eq. \eqref{equ:n_bt_lower_bound}.

\section{Coin acceptance probabilities for all IBMQ platforms}
\label{sec:coin_acceptance_all}

\begin{table*}[t!]
\begin{tabular}{|l|r||r||c|c|c|c|}
	\hline
	IBMQ & $M$ & $n_\text{cT}$ & $\overline{p}_{\text{Mf}_0}$ & $\overline{p}_{\text{Mf}_1}^{(\text{opt})}$ & $\overline{p}_{\text{Mf}_2}^{(\text{opt})}$ & $\overline{p}_{\text{Mf}_3}^{(\text{fix})}$ \\ 
	\hline
	          &     1 &     1 & 0.03226 & 0.1409 & 0.3196 & 0.4832 \\
	          &     9 &     9 & 3.784$\cdot10^{-14}$ & 2.189$\cdot10^{-8}$ & 3.479$\cdot10^{-5}$ & 0.001436 \\
	          &    16 &    16 & 1.376$\cdot10^{-24}$ & 2.413$\cdot10^{-14}$ & 1.185$\cdot10^{-8}$ & 8.831$\cdot10^{-6}$ \\
	          &    25 &    25 & 5.207$\cdot10^{-38}$ & 5.282$\cdot10^{-22}$ & 4.123$\cdot10^{-13}$ & 1.268$\cdot10^{-8}$ \\
	Sherbrooke&    64 &    64 & 3.585$\cdot10^{-96}$ & 3.391$\cdot10^{-55}$ & 1.972$\cdot10^{-32}$ & 6.083$\cdot10^{-21}$ \\
              &   100 &   100 & 7.353$\cdot10^{-150}$ & 7.782$\cdot10^{-86}$ & 2.889$\cdot10^{-50}$ & 2.586$\cdot10^{-32}$ \\
	          &  1024 &  1024 & 7.453$\cdot10^{-1528}$ & 3.054$\cdot10^{-872}$ & 5.216$\cdot10^{-508}$ & 3.515$\cdot10^{-324}$ \\
	          & 10000 & 10000 & 4.402$\cdot10^{-14914}$ & 1.288$\cdot10^{-8511}$ & 1.169$\cdot10^{-4954}$ & 1.859$\cdot10^{-3159}$ \\
	\hline
              &     1 &     1 & 3.370$\cdot10^{-2}$ & 0.1408 & 0.3099 & 0.4651 \\
	          &     9 &     9 & 5.606$\cdot10^{-14}$ & 2.175$\cdot10^{-8}$ & 2.636$\cdot10^{-5}$ & 0.001018 \\
	          &    16 &    16 & 2.767$\cdot10^{-24}$ & 2.386$\cdot10^{-14}$ & 7.237$\cdot10^{-9}$ & 4.794$\cdot10^{-6}$ \\
	          &    25 &    25 & 1.551$\cdot10^{-37}$ & 5.189$\cdot10^{-22}$ & 1.908$\cdot10^{-13}$ & 4.883$\cdot10^{-9}$ \\
	  Kyiv    &    64 &    64 & 5.866$\cdot10^{-95}$ & 3.240$\cdot10^{-55}$ & 2.743$\cdot10^{-33}$ & 5.284$\cdot10^{-22}$ \\
              &   100 &   100 & 5.794$\cdot10^{-148}$ & 7.249$\cdot10^{-86}$ & 1.325$\cdot10^{-51}$ & 5.683$\cdot10^{-34}$ \\
	          &  1024 &  1024 & 1.963$\cdot 10^{-484}$ & 1.476$\cdot 10^{-872}$ & 1.025$\cdot 10^{-521}$ & 3.694$\cdot 10^{-341}$ \\
	          & 10000 & 10000 & 1.991$\cdot 10^{-14724}$ & 1.063$\cdot 10^{-8514}$ & 1.643$\cdot 10^{-5088}$ & 2.907$\cdot 10^{-3325}$ \\
    \hline
	          &     1 &     1 & 0.1836 & 0.4726 & 0.7372 & 0.9384 \\
	          &     9 &     8 & 9.724$\cdot10^{-6}$ & 0.01299 & 0.2706 & 0.8976 \\
	          &    16 &    15 & 1.203$\cdot10^{-10}$ & 1.168$\cdot10^{-4}$ & 5.101$\cdot10^{-2}$ & 0.7415 \\
	          &    25 &    23 & 2.389$\cdot10^{-15}$ & 2.932$\cdot10^{-6}$ & 2.351$\cdot10^{-2}$ & 0.8026 \\
	Osaka     &    64 &    62 & 3.101$\cdot10^{-43}$ & 3.800$\cdot10^{-18}$ & 9.389$\cdot10^{-7}$ & 0.2374 \\
	          &   100 &    98 & 2.398$\cdot10^{-69}$ & 1.767$\cdot10^{-29}$ & 3.818$\cdot10^{-11}$ & 0.05007 \\
	          &  1024 &  1020 & 2.861$\cdot 10^{-741}$ & 3.405$\cdot 10^{-323}$ & 1.900$\cdot 10^{-127}$ & 4.773$\cdot10^{-23}$ \\
	          & 10000 &  9990 & 4.420$\cdot 10^{-7322}$ & 7.117$\cdot 10^{-3222}$ & 6.504$\cdot 10^{-1296}$ & 3.140$\cdot 10^{-255}$ \\
	\hline
	          &     1 &     1 & 0.3045 & 0.6475 & 0.8537 & 0.9776 \\
	          &     9 &     8 & 4.851$\cdot10^{-4}$ & 0.1180 & 0.6123 & 0.9837 \\
	          &    16 &    15 & 2.051$\cdot10^{-7}$ & 9.270$\cdot10^{-3}$ & 0.2978 & 0.9511 \\
	          &    25 &    24 & 7.143$\cdot10^{-12}$ & 2.790$\cdot10^{-4}$ & 0.1013 & 0.8927 \\
	Brisbane  &    64 &    62 & 9.496$\cdot10^{-30}$ & 5.260$\cdot10^{-10}$ & 2.857$\cdot10^{-3}$ & 0.8269 \\
	          &   100 &    98 & 5.951$\cdot10^{-48}$ & 2.025$\cdot10^{-16}$ & 2.208$\cdot10^{-5}$ & 0.6113 \\
	          &  1024 &  1021 & 3.323$\cdot 10^{-520}$ & 1.483$\cdot 10^{-186}$ & 4.143$\cdot10^{-65}$ & 2.058$\cdot10^{-7}$ \\
	          & 10000 &  9993 & 4.794$\cdot 10^{-5138}$ & 7.021$\cdot 10^{-1866}$ & 9.746$\cdot 10^{-669}$ & 2.771$\cdot10^{-86}$ \\
	 \hline
	          &     1 &     1 & 0.5128 & 0.8367 & 0.9192 & 0.9583 \\
	          &     9 &     8 & 0.02342 & 0.5540 & 0.8391 & 0.9485 \\
	          &    16 &    15 & 3.704$\cdot10^{-4}$ & 0.2378 & 0.6251 & 0.8580 \\
	          &    25 &    24 & 1.388$\cdot10^{-6}$ & 6.817$\cdot10^{-2}$ & 0.3891 & 0.7198 \\
	Kyoto     &    64 &    62 & 5.143$\cdot10^{-16}$ & 1.000$\cdot10^{-3}$ & 0.1011 & 0.4978 \\
	          &   100 &    98 & 4.510$\cdot10^{-26}$ & 3.778$\cdot10^{-6}$ & 0.01053 & 0.2081 \\
	          &  1024 &  1020 & 3.614$\cdot 10^{-287}$ & 3.470$\cdot10^{-72}$ & 9.670$\cdot10^{-32}$ & 2.044$\cdot10^{-14}$ \\
	          & 10000 &  9991 & 5.237$\cdot 10^{-2871}$ & 5.651$\cdot 10^{-751}$ & 1.095$\cdot 10^{-345}$ & 1.623$\cdot 10^{-169}$ \\
	 \hline
\end{tabular}
\caption{Acceptance probability for forged coins for all IBMQ platforms containing 1, 9, 100, 1024, 10000 quantum tokens, each one with $N=30$ qubits. Here, the acceptance probability of the bank coins with more than 1 token is always bigger than 0.99999, where we take 100 digits of precision.}
\label{tab:hack_coin_all}
\end{table*}

\section{Glossary of main variables}
\label{sec:main_variables}

\begin{table*}[t!]
	\begin{tabular}{|c c|}
		\hline
		Variable & Physical quantity \\ \hline
		$\theta$ & polar angle on the Bloch sphere \\
		$\phi$ & azimuthal angle on the Bloch sphere \\

		$\theta_\text{b}, \phi_\text{b}$ & angles which the bank prepares and measures the token \\ 
		$\theta_{\text{f}_j}, \phi_{\text{f}_j}$ & angles used by the attacker to measure the bank token in the $j$-th measurement \\
		$\theta_\text{f}, \phi_\text{f}$ & angles forged by the attacker \\
		
		$n_{\text{f}_j}$ & number of photons measured in the $j$-th measurement \\
		$N_j$ & number of qubits used in the $j$-th measurement\\
		
		$N$ & total number of qubits in the quantum token \\
		$P_0$ & probability to detect a photon if qubit is in state $|0\rangle $\\
		$P_1$ & probability to detect a photon if qubit is in state $|1\rangle $ \\
		$\sigma_N$ & total uncertainty of photon counts \\
		$\overline{n}$ & averaged normalized counts of photons \\
		
		$p_\text{q}$ & probability that a qubit emits a photon \\
		$p_\text{t}$ & probability that a quantum token emits a given number of photons\\
		
		$\overline{p}_{\text{b}}$ & average probability of acceptance for bank token\\
		
		$\overline{p}_{\text{f}_j}$ & average probability of acceptance for forged token generated from $j$-measurements\\
		
		$\varepsilon_\text{b}$ & limit for the bank declines own token\\
		$\varepsilon_\text{cb}$ & limit for bank declines own coins\\
		$\varepsilon_\text{cf}$ & limit for bank accepting forged coins\\
		
		$n_\text{T}$ & photon count threshold for accepting the quantum token  \\
		$n_\text{cT}$ & minimum number of accepted token for accepting coin \\
		$M$ & number of quantum tokens in the coin \\
		$p_\text{b}$ & self-acceptance probability of the bank tokens \\
		$p_\text{f}$ & acceptance probability of forged tokens \\
		$\mathcal{L}$ & likelihood function\\
		\hline
	\end{tabular}
	\caption{Glossary of the main variables in the text.}
\label{tab:variables}
\end{table*}


%



\end{document}